\documentclass[twocolumn]{aastex631}
\usepackage[utf8]{inputenc}
\usepackage[T1]{fontenc}
\usepackage{mathptmx}
\usepackage{ulem}
\usepackage{etoolbox}
\usepackage{graphicx}

\usepackage{appendix}
\usepackage{booktabs}

\usepackage{amsmath}
\usepackage{braket}
\usepackage{color}

\begin{document}

\title{Miscibility and Transport Properties in Hydrogen–Neon Mixtures}

\author{Armin Bergermann}
\author{Siegfried Glenzer}%
\author{Arianna Gleason}%
\affiliation{SLAC National Accelerator Laboratory, Menlo Park CA 94309, USA}%
\author{Ronald Redmer}
\affiliation{Institute of Physics, University of Rostock, D-18051 Rostock, Germany}
\affiliation{Helmholtz-Zentrum Dresden-Rossendorf, Bautzner Landstr. 400, D-01328 Dresden, Germany}%

\date{\today}

\begin{abstract}
The mixing behavior of hydrogen with heavier elements plays a key role in modeling the interiors of giant planets such as Jupiter and Saturn. Using density functional theory combined with molecular dynamics, we investigate hydrogen–neon mixtures and find that the minimum pressure required to trigger phase separation is substantially lower than in hydrogen–helium mixtures. Our simulations further reveal that the presence of neon stabilizes hydrogen molecules even at temperatures of $\approx 10000$~K and pressures of $\approx 10$~Mbar, similar to trends observed in hydrogen–helium mixtures but significantly more pronounced. This stabilization is accompanied by a reduction of several orders of magnitude in the electrical conductivity compared to pure hydrogen. These results, together with the larger X-ray scattering cross section of neon, establish hydrogen–neon as a valuable experimental surrogate for probing phase separation in hydrogen-rich mixtures and provide new insight into the physical mechanisms in hydrogen and mixtures with heavier elements under planetary interior conditions.
\end{abstract}


\section{Introduction}
The miscibility gap of hydrogen (H) with heavier elements such as helium (He)~(\cite{Brygoo2021, Helled2020b, Schoettler2018, Morales2009, Bergermann2021a, Lorenzen2009, Pfaffenzeller1995, Stevenson1977a, Chang2024, Nettelmann2024}), carbon~(\cite{Cheng2023, Militzer2024}) and water~(\cite{Bergermann2021b, Bergermann2024, Gupta2025, Soubiran2015b}) under extreme $p$-$T$ conditions plays a crucial role in planetary modeling~(\cite{Militzer2024, Howard2024, Howard2025, Mankovich2020}). Phase separation in such mixtures can significantly impact the thermal evolution and internal structure of gas and ice giant planets, and is essential to interpret data from both solar system missions and exoplanetary observations~(\cite{Howard2024, Mankovich2020, Bailey2021, Scheibe2021, Helled2020a, Nettelmann2016, Marley1995, Masters2014, Arridge2014, Stevenson1977a, Howard2025, Cano2024, Rauer2014}).

Among these mixtures, H–He has received particular attention. First-principles simulations focused on the investigation of the miscibility gap~(\cite{Schoettler2018, Morales2013b, Lorenzen2009, Karasiev2026}). The findings indicate that H-He phase separates under conditions relevant to the outer regions of Jupiter, Saturn, and gas giant exoplanets. The separation process leads to the formation of He-rich droplets sinking toward the planetary core, increasing the planet's internal heat budget~(\cite{Howard2024, Mankovich2020, Nettelmann2013a, Stevenson1977a, Lorenzen2009, Morales2009}). Note that earlier studies have identified the metallization of H as a catalyst for phase separation~(\cite{Lorenzen2009, Lorenzen2011, Schoettler2018, Knudson2015}). To date, only one experimental campaign has been reported under $p$-$T$ conditions relevant for the interior of giant gas planets~(\cite{Brygoo2021}). Unfortunately, the experimental measurements largely disagree with theoretical predictions~(\cite{Brygoo2021, Chang2024, Schoettler2018}). The main experimental challenge lies in the low X-ray scattering contrast of H and He and the difficulties in preparing H-He mixtures under well-controlled $p$-$T$ conditions. 

To address these experimental limitations and further improve our understanding of H mixtures with other noble gases under relevant $p$–$T$ conditions for gas giant planets such as Jupiter and Saturn, we turn to H–Neon (Ne) mixtures as an experimentally accessible surrogate system. Ne is chemically inert and non-bonding, much like He, but offers significantly improved experimental accessibility because of its higher atomic number and X-ray contrast. Both H and Ne have been extensively characterized as pure substances under extreme conditions, see e.g.~(\cite{Bonitz2024, Nghia2022, Wang2024, Driver2015, Vos1991, McWilliams2015, Knudson2001, Knudson2015, Ross1983, Ross1996}). For H, particular attention has been given to the first-order liquid–liquid transition accompanied by a non-metal-to-metal transition. This phenomenon has been investigated using a variety of \textit{ab initio} approaches~(\cite{Bergermann2024b, Lorenzen2010, Hinz2020, Scandolo2003, Morales2010a, Geng2019, Pierleoni2016, Mazzola2018, Bund2021}) as well as experimental studies~(\cite{Zaghoo2016, Ohta2015, Dzyabura2013, Celliers2018, Knudson2015}). Likewise, Ne’s equation of state (EOS)~\cite{Driver2015, Militzer2021} and melting line~(\cite{Koci2007a, Nghia2022, Vos1991, Tang2017, Wang2024, He2010}) had been investigated in detail.

Wilson and Militzer~(\cite{Wilson2010}) explored the solubility of Ne and Ar in H-He mixtures, predicting that Ne prefers the He-rich phase, while Ar is excluded. Despite this foundation, the phase behavior of H–Ne mixtures remains largely unexplored.

The key thermodynamic potential that governs phase separation in mixtures is the Gibbs free energy
\begin{equation}
G(T, p, N) = U - TS + pV\textrm{,}
\end{equation}
where $T$ is the temperature, $p$ the pressure, $S$ the entropy, and $V$ the volume. The stability of a binary mixture can be assessed by the Gibbs free energy of mixing
\begin{equation}
\Delta G(p,T,x_{\textrm{H}}) = G(p,T,x_{\textrm{H}}) - x_{\textrm{H}} G(p,T,1) - (1 - x_{\textrm{H}}) G(p,T,0)\textrm{,}
\end{equation}
where $x_{\textrm{H}}$ is the H concentration and $G(p,T,x)$ denotes the Gibbs free energy of the mixture (with pure H and pure Ne corresponding to $x=1$ and $x=0$, respectively). A concave region in $\Delta G$ indicates thermodynamic instability and spontaneous phase separation into two co-existing phases with different compositions~(\cite{Oliveira2017}).

To determine miscibility gaps from \textit{ab initio} simulations, a common approach involves evaluating $\Delta G$, using density functional theory and molecular dynamics (DFT-MD) simulations combined with corrections for nonideal entropy contributions~(\cite{Schoettler2018, Morales2013b, Soubiran2015b, Bergermann2024}). First, the EOS is calculated. Then, the deviations from the ideal mixing entropy are calculated by coupling constant integration and thermodynamic integration (TI)~(\cite{Kirkwood1935}), allowing for a precise construction of the miscibility diagram. To avoid spontaneous phase separation within the simulation cell, these calculations typically employ small system sizes-often limited to a few dozen electrons. This approach was successfully applied to H–He mixtures~(\cite{Schoettler2018, Morales2013}) but remains limited in scope, as it does not provide direct access to structural properties. Additionally, the small system size might lead to finite-size effects, e.g. deviations in pressure, energy, and structural properties.

Recent developments in machine-learned interatomic potentials offer promising alternatives. These models aim to combine the precision of \textit{ab initio} simulations with the computational efficiency required for large-scale MD simulations, allowing direct exploration of phase separation and interfacial structures~(\cite{Cheng2023, Chang2024}). Furthermore, structural metrics, particularly the height of the first peak in the pair correlation function (PDF), provide a practical approach to estimate the $p$ - $T$ conditions where the system transitions from fully mixed to phase separated~(\cite{Karasiev2026}). 

In this work, we use DFT-MD to investigate H–Ne mixtures under extreme $p$--$T$ conditions. We identify the onset of phase separation through structural signatures in the PDFs. Next, we examine the stabilization of H$_2$ molecules in the presence of Ne through a bond-lifetime analysis. In addition, we compute diffusion coefficients and electronic transport properties to explore how H metallization influences phase separation and transport behavior. In the Supplementary Material (SM), a tabulated EOS is provided for H-Ne mixtures for all conditions investigated in this work which could be used to interperete future experimental campaigns.

\section{Methods} We used the Born-Oppenheimer approximation to decouple electronic and ionic degrees of freedom. The electronic subsystem was treated using DFT, while the ionic positions were evolved in time using classical MD, as implemented in the Vienna \textit{Ab initio} Simulation Package (VASP)~(\cite{Kresse1993, Kresse1994, Kresse1996b, Kresse1996a}). We used the Perdew–Burke–Ernzerhof (PBE) exchange-correlation (XC) functional,~(\cite{Perdew1996}) a plane-wave energy cut-off of $1000$~eV and the Baldereschi mean value point to sample the Brillouin zone. Projector-augmented wave (PAW) pseudopotentials were used: PAW\_PBE H\_h 06Feb2004 for H and PAW\_PBE Ne 05Jan2001 for Ne.

MD simulations employed a time step of $0.3$~fs and total simulation times of up to $40$~ps to ensure statistical convergence and to reduce fluctuations in the calculated observables. Following an initial equilibration phase (typically $5000$ timesteps), we computed observables averaged over time, specifically pressure and total energy, for a range of H concentrations $x_{\textrm{H}}=0.125,0.250,0.375,0.500,0.625,0.750,\textrm{ and }0.875$, where $x_{\textrm{H}}=n_{\textrm{H}}/(n_{\textrm{H}}+n_{\textrm{Ne}})$ with $n_{\textrm{H}}$ and $n_{\textrm{Ne}}$ being the number of H and Ne atoms, respectively. Although pressure and energy already converge with $128$ atoms, simulations with $256$ atoms were required to obtain converged structural properties. Detailed convergence tests with $128$, $256$, and $512$ atoms are presented in the SM.

Unlike the H–He system, where small simulation cells can suppress phase separation,~(\cite{Schoettler2018}) we observed spontaneous phase separation in H–Ne mixtures even for small systems (e.g., $16$ atoms). This complicates the application of thermodynamic integration (TI), which requires a reversible path within a single homogeneous phase and therefore cannot be reliably performed across phase transitions~(\cite{Pin2023}). To overcome this limitation, we adopted the approach proposed by Karasiev et al.~(\cite{Karasiev2026}), which estimates the onset of miscibility from structural signatures. Specifically, we monitored the pressure dependence of the height of the first peak in the H--Ne PDF along isotherms to determine the highest pressure at which the mixed phase remains stable. In the phase-separated regime, the formation of an interface between H and Ne reduces the probability of neighbors of a different type, thereby suppressing the first peak in the H-Ne PDF. Upon decreasing pressure, the system transitions to a mixed state and the peak height increases, reaching a maximum or plateau at the miscibility threshold. At still lower pressures, the peak height declines again as interactions weaken in the more dilute fluid. Earlier studies inferred miscibility solely from the long-range behavior of the PDFs~(\cite{Lorenzen2011}); we provide a direct comparison with this approach in the SM. In addition, we visually inspect all simulations cells to further verify our findings and to avoid misinterpretation.

Bond lifetimes were evaluated by tracking H--H pairs along the MD trajectories. Two H atoms were classified as bonded whenever their instantaneous separation fell below $1.0$~Å, the approximate first minimum in the H–H PDF. Tracking these bonded pairs across consecutive MD frames yields a set of discrete lifetimes, each corresponding to one continuous interval during which a given H–H pair remained within the bonding distance. These lifetimes were then accumulated into a histogram, where the x-axis represents the lifetime in femtoseconds and the y-axis gives the number of distinct H$_2$ bonding episodes with lifetimes falling into each bin.

Furthermore, we evaluated the electronic transport properties of the H-Ne mixtures using the Kubo–Greenwood formalism, as implemented in VASP (see Refs.~\cite{French2017, Gajdos2006, Holst2011, Knyazev2013}). For each $\rho$–$T$ state point, we extracted $20$ statistically independent snapshots and computed conductivities using the PBE XC functional together with a $2\times2\times2$ Monkhorst–Pack~(\cite{Monkhorst1976}) \textbf{k}-point mesh. These settings ensure converged results for the conductivity across the entire range of conditions studied.

\begin{figure}
\centering
\includegraphics[width=\hsize]{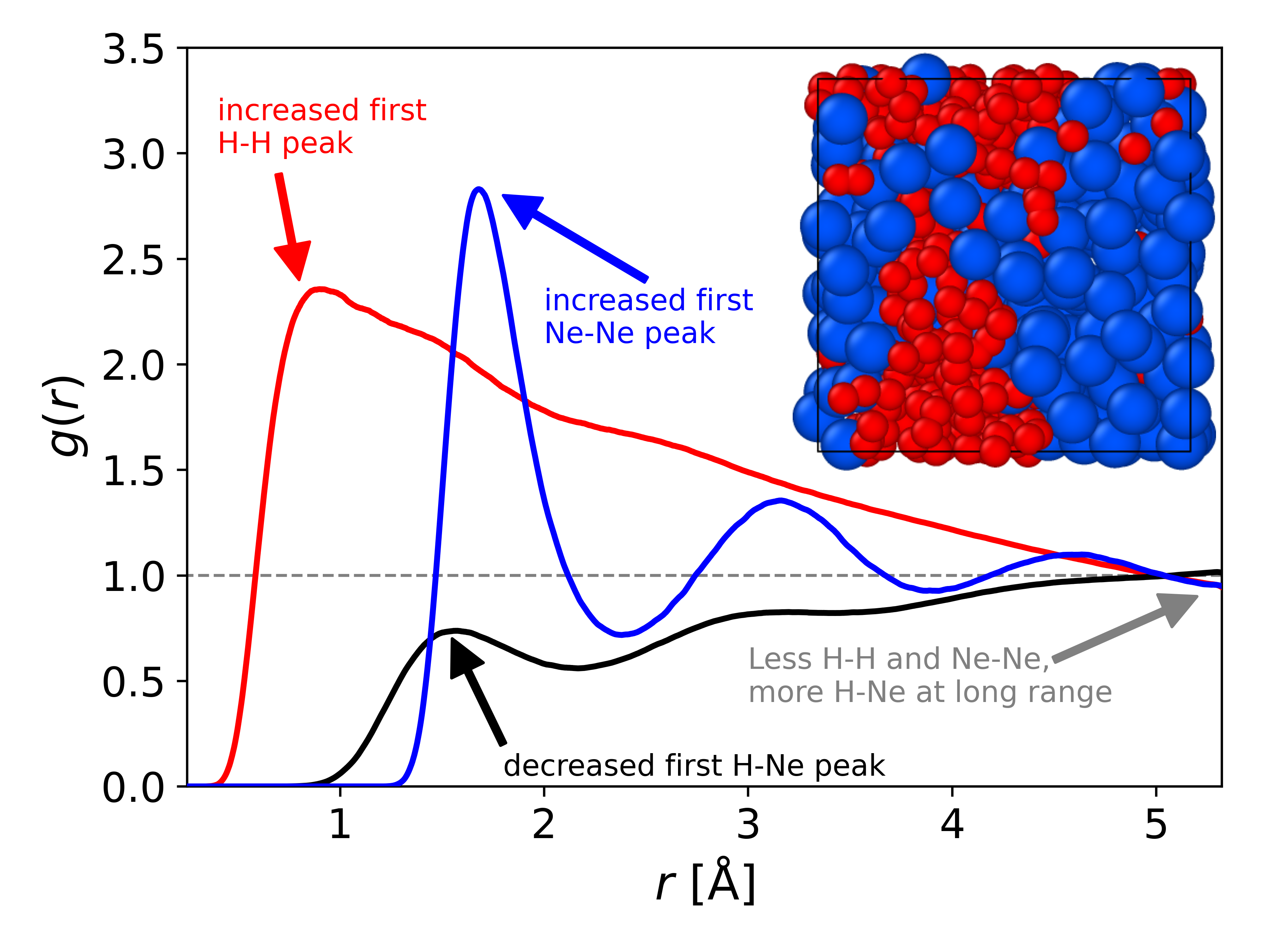}
	\caption{Pair distribution function of H-H (red), H-Ne (black), and Ne-Ne (blue) for a temperature of $10000$~K and a pressure of $11.5$~Mbar. The inset shows a typical snapshot of the $320$ H (red) and $192$ Ne (blue) atoms.}
    \label{SimCell}
\end{figure}

\section{Results}
To characterize the microscopic structure and ionic transport properties of H--Ne mixtures, we analyze pair distribution functions (Sec.~\ref{sec:PDF}), molecular bond lifetimes (Sec.~\ref{sec:BL}), and self-diffusion coefficients (Sec.~\ref{sec:DC}) obtained from our DFT-MD simulations. Next, we discuss the miscibility gap of H--Ne (Sec.~\ref{sec:MG}) and compare our results with H--He mixtures. Finally, we analyze the electrical and thermal conductivities (Sec.~\ref{sec:KG}).

\subsection{Pair distribution functions}\label{sec:PDF}
Inspecting the PDFs (Fig.~\ref{SimCell}) of a DFT-MD simulation at $10000$~K and $11.5$~Mbar for a H concentration of $x_{\mathrm{H}} = 0.625$ clearly reveals phase separation within the simulation cell. For H–Ne, the probability of finding atoms in close proximity is reduced, while it increases at larger separations $(>5$~\AA$)$. In contrast, H–H and Ne–Ne PDFs exhibit enhanced probability at short distances but reduced probability at large distances. This spatial correlation pattern is indicative of phase separation~(\cite{Lorenzen2009, Karasiev2026}), with coexisting regions enriched in H or Ne. A representative snapshot of the simulation cell is shown alongside the PDFs, which visually confirms the phase-separated structure. We emphasize that the characteristic size of the phase-separated domains is limited by the finite particle number, and thus does not reflect the macroscopic droplet size expected in the thermodynamic limit.

\begin{figure}
\centering
\includegraphics[width=\hsize]{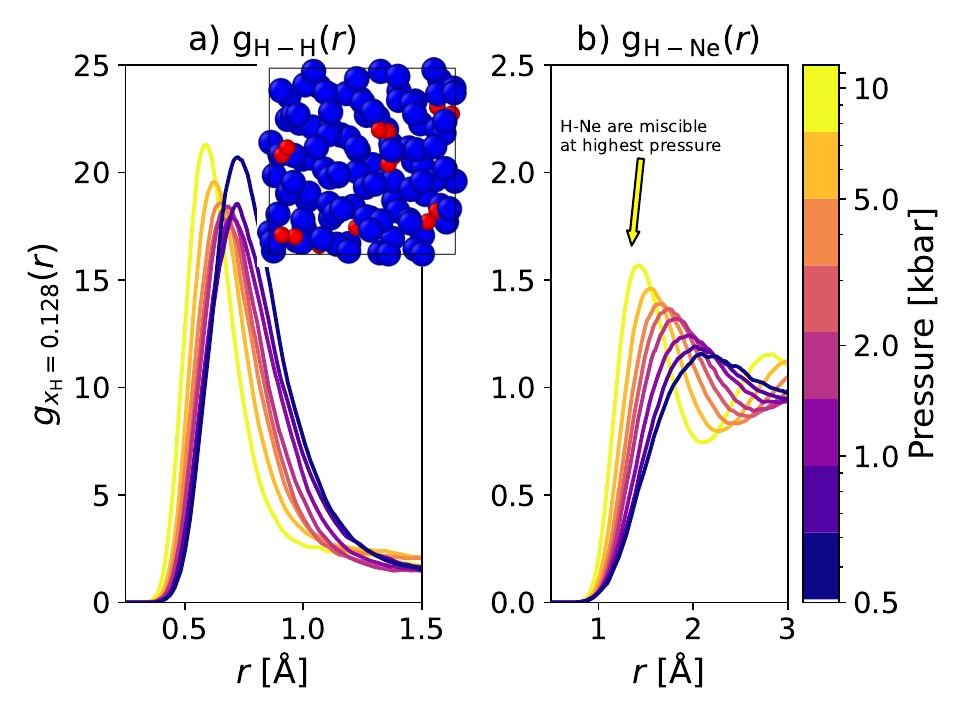}
	\caption{Pair distribution functions $g_{\textrm{H-Ne}}(r)$ for $x_{\textrm{H}}=0.128$, various pressures (color-coded) and a temperature of $10000$~K. Panel a) shows the H-H and b) shows the  H-Ne PDF. Additionally, we present a typical snapshot of the simulation cell as an inset.}
    \label{PDF0128}
\end{figure}

\begin{figure}
\centering
\includegraphics[width=\hsize]{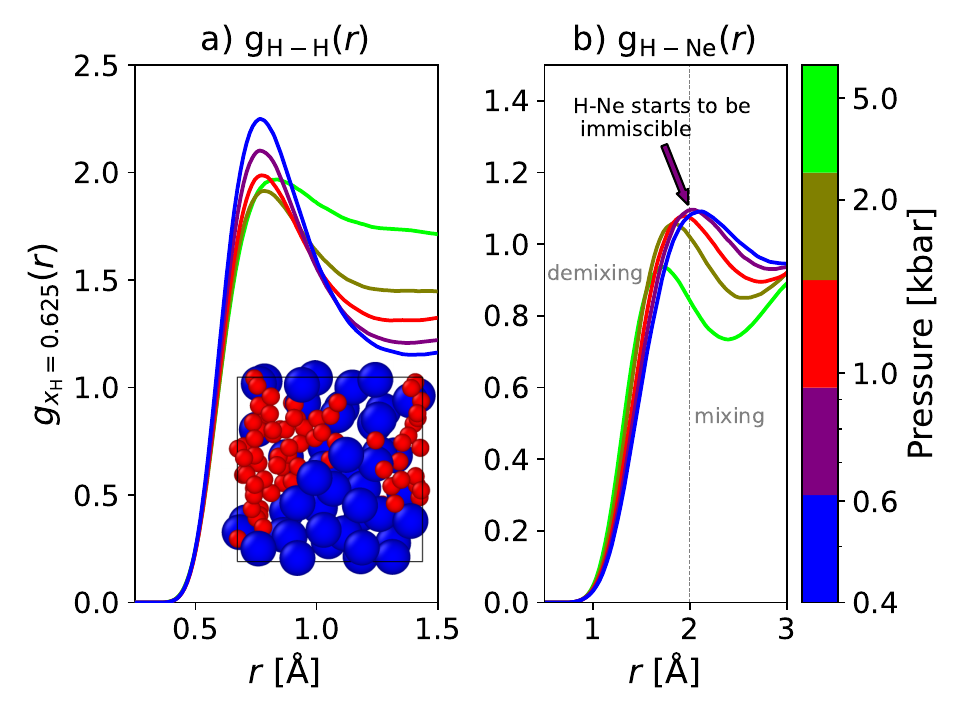}
	\caption{Pair distribution functions $g_{\textrm{H-Ne}}(r)$ for $x_{\textrm{H}}=0.625$, various pressures (color-coded) and a temperature of $10000$~K. Panel a) shows the H-H and b) shows the  H-Ne PDF. Additionally, we present a typical snapshot of the simulation cell as an inset.}
    \label{PDF0625}
\end{figure}

We analyze the pressure dependence in the height in the first peak in the H–Ne PDFs to estimate the $p$-$T$ conditions for the onset of phase separation. First, we analyze the H--H and H--Ne PDFs for a Ne-rich mixture with $x_{\mathrm{H}} = 0.128$ at $10000$~K (Fig.~\ref{PDF0128}a). Distinct H$_2$ molecular peaks persist even at $p$--$T$ conditions far above the well-known liquid–liquid transition for pure H~(\cite{Bergermann2024b, Lorenzen2010, Hinz2020, Scandolo2003, Morales2010a, Geng2019, Pierleoni2016, Mazzola2018, Bund2021}), indicating that Ne strongly stabilizes the molecular bonding of H$_2$. This finding parallels previous results in H–He mixtures,~(\cite{Vorberger2007}) but the effect is much more pronounced for H–Ne: the large Ne atoms occupy substantial volume, restricting the available configurational space for H and thereby suppressing electronic delocalization and metallization. With increasing pressure, the H–H bond length decreases further; at $16$~Mbar, we find an average value of $0.6$~Å, significantly shorter than the equilibrium bond length of an isolated H$_2$ molecule ($\approx 0.74$~Å)~(\cite{Syrkin1964}).

Additionally, we present a typical snapshot of our simulation cell at a pressure of $\approx 15$~Mbar (inset in Fig~\ref{PDF0128}a), showing abundant H$_2$ molecules but no indication of phase separation at this low concentration of H. Fig~\ref{PDF0128}b displays the corresponding H–Ne PDFs. The monotonic increase in the height of the first peak with pressure demonstrates that H and Ne remain fully miscible in this regime; a maximum and subsequent decrease would signal the onset of phase separation, which is not observed here.

The H-H PDFs for a more H-rich mixture ($x_{\mathrm{H}} = 0.625$) at the same temperature (Fig.~\ref{PDF0625}a) exhibit only a weak molecular peak consistent with a largely dissociated fluid containing transient H$_2$ bonds. A snapshot at $5$~Mbar illustrates that this composition does undergo phase separation. The height of the first peak in the H-Ne PDF (Fig.~\ref{PDF0625}b) increases with pressure and reaches a maximum at $0.93$~Mbar, beyond which it drops sharply—signaling the emergence of phase separation at higher pressures. The long-range features and the peak heights for all calculated $p$--$T$ conditions are shown in the SM.

\begin{figure}
\centering
\includegraphics[width=\hsize]{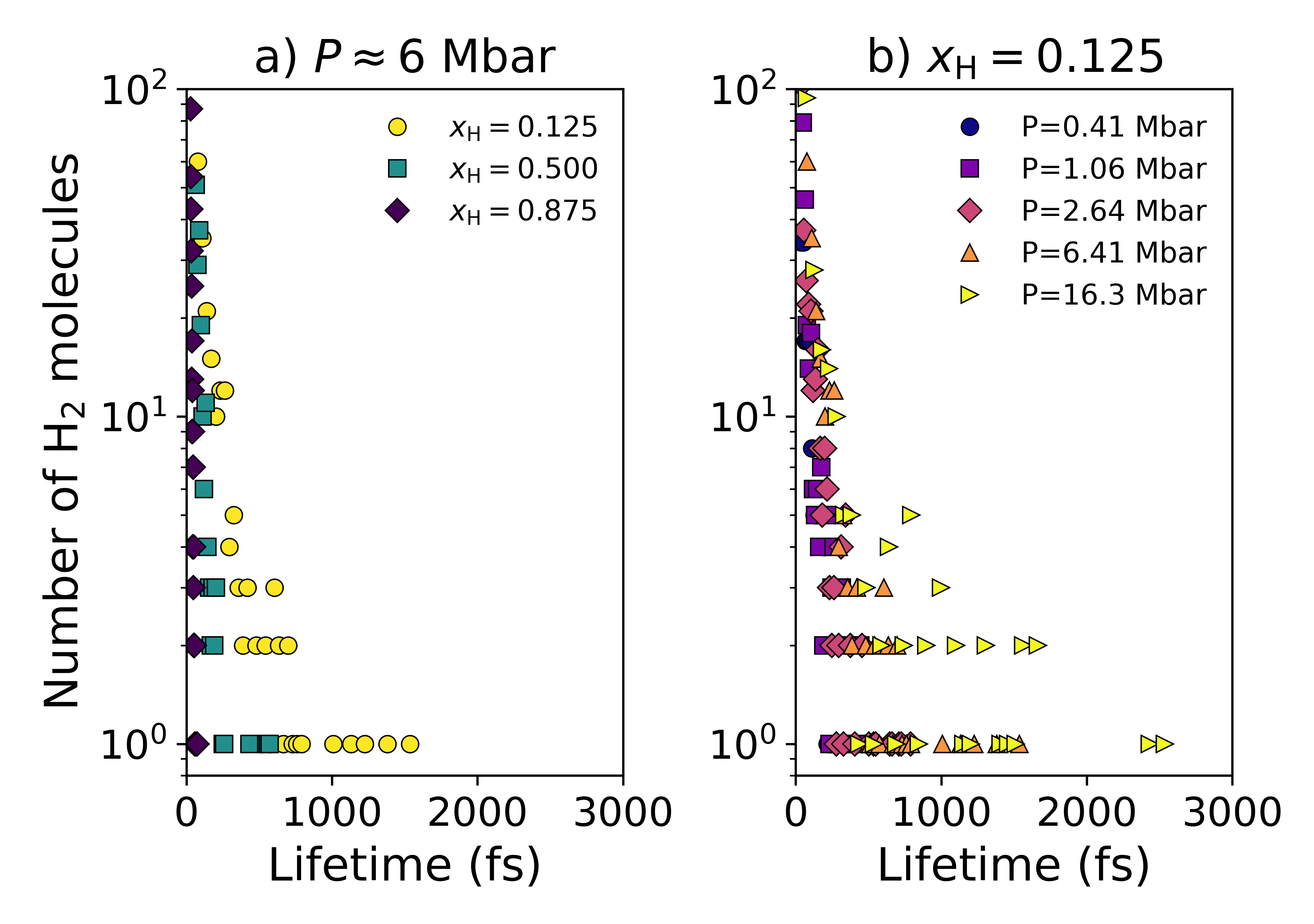}
	\caption{Lifetimes of H$_2$ molecules at $10000$~K. Panel a) shows results for different concentrations and a pressure of $\approx 6$~Mbar. Panel b) depicts the lifetimes of H$_2$ molecules for a H concentration of x$_{\textrm{H}}=0.125$ and different pressures (color-coded).}
    \label{Lifetimes}
\end{figure}

\subsection{Bond-lifetime analysis}\label{sec:BL}
To further quantify the behavior of H$_2$ molecules and distinguish persistent bonding from transient interactions, we analyze the distribution of H--H bond lifetimes. This dynamical measure complements the static PDFs: while the PDFs capture average structural correlations, the lifetime distribution directly reveals the relative abundance of short-lived versus longer-living H$_2$ molecules and thereby provides insight into the dynamical stability of molecular H$_2$ in the mixture.

First, we analyze the results for $10000$~K and $\approx 6$~Mbar for three H compositions $x_{\mathrm{H}} = 0.125, 0.5, 0.875$ (Fig.~\ref{Lifetimes}a). The most Ne-rich mixture exhibits the longest-lived H$_2$ molecules, with lifetimes reaching up to $\sim 1500$~fs. This behavior confirms that Ne suppresses dissociation and stabilizes molecular H$_2$ at low H concentrations, consistent with trends previously reported for H–He mixtures.~(\cite{Vorberger2007, Lorenzen2011}) Additionally, we investigate the dependence of the lifetimes for a fixed composition of $x_{\mathrm{H}} = 0.875$ (Fig.~\ref{Lifetimes}b). Increasing pressure leads to a systematic increase in molecular lifetime, consistent with the progressive compression of H atoms and the stabilization of a more localized, H$_2$-like electronic structure. Consequently, dissociation events become rarer, and the surviving H$_2$ pairs persist for longer durations.

A convergence test with respect to the bonding cutoff confirmed that the absolute lifetimes of individual H$_2$ molecules vary slightly with the chosen threshold but qualitative trends remain robust.

\subsection{Self-diffusion coefficients}\label{sec:DC}
\begin{figure}
\centering
\includegraphics[width=\hsize]{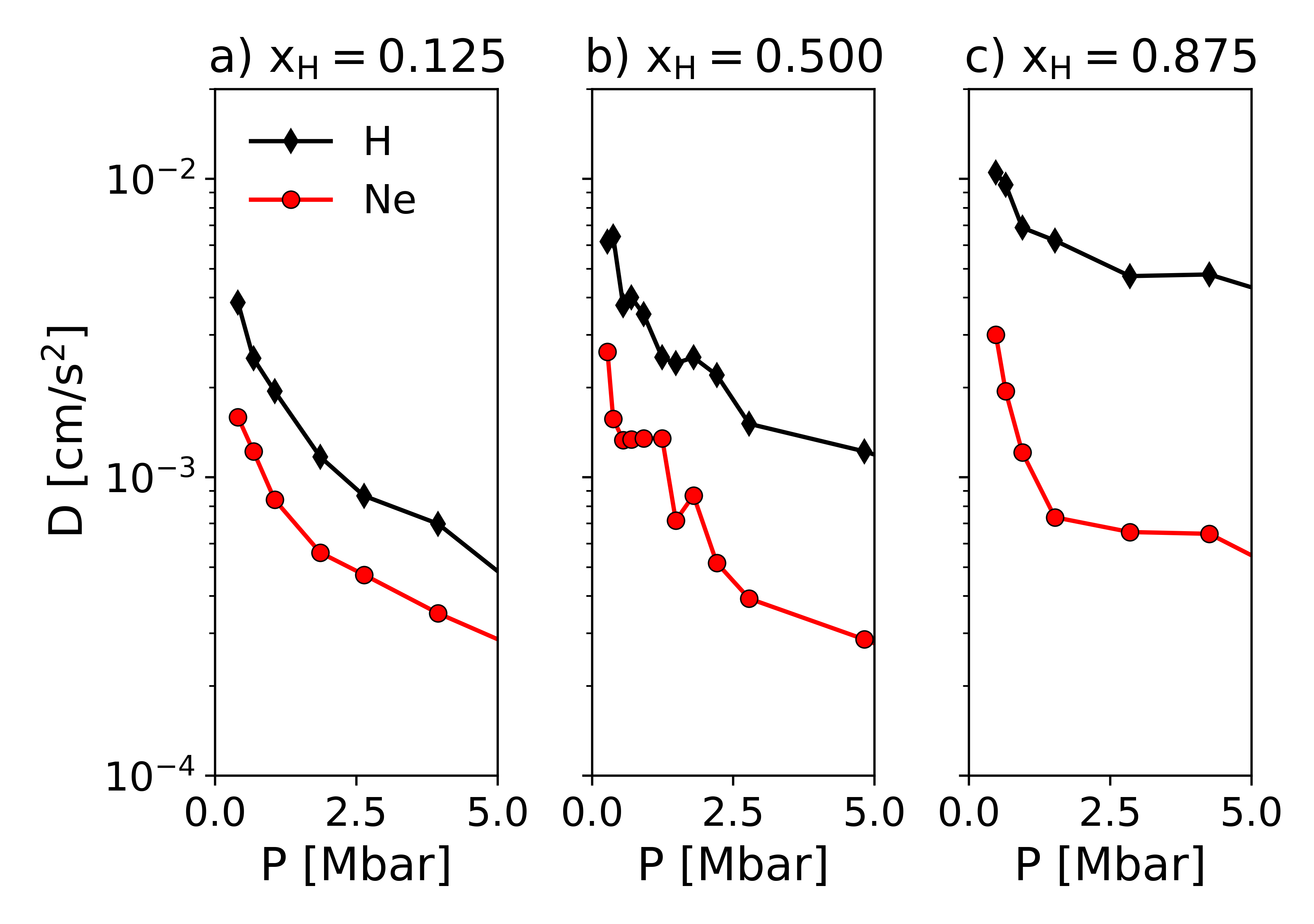}
	\caption{Self-diffusion coefficients of H and Ne at $T=10{,}000$~K as a function of pressure for mixtures with $x_\mathrm{H}=0.125$, $0.500$, and $0.875$, obtained from mean-square displacement analysis.}
    \label{fig:Diffusion}
\end{figure}

We present self-diffusion coefficients for $10000$~K (Fig.~\ref{fig:Diffusion}) and H concentrations of  $x_\mathrm{H}=0.125$, $0.5$, and $0.875$, obtained from the slope of the mean-square displacement. Within the statistical uncertainties, the Ne self-diffusion coefficient shows only a weak dependence on composition and remains comparable across all three mixtures. In contrast, the H self-diffusion is strongly reduced at low H concentrations, decreasing by up to an order of magnitude in Ne-rich mixtures.

This trend is consistent with the structural analysis discussed above: increasing the Ne fraction reduces the phase space available to H and enhances its molecular character. As a result, the mobility of H becomes increasingly suppressed in Ne-rich mixtures, while the transport of Ne remains comparatively insensitive to composition. Unfortunalety, statistical fluctuations and finite-size effects introduce significant uncertainty in the diffusion coefficients, limiting their usefulness for identifying the onset of phase separation.

\begin{figure}
\centering
\includegraphics[width=\hsize]{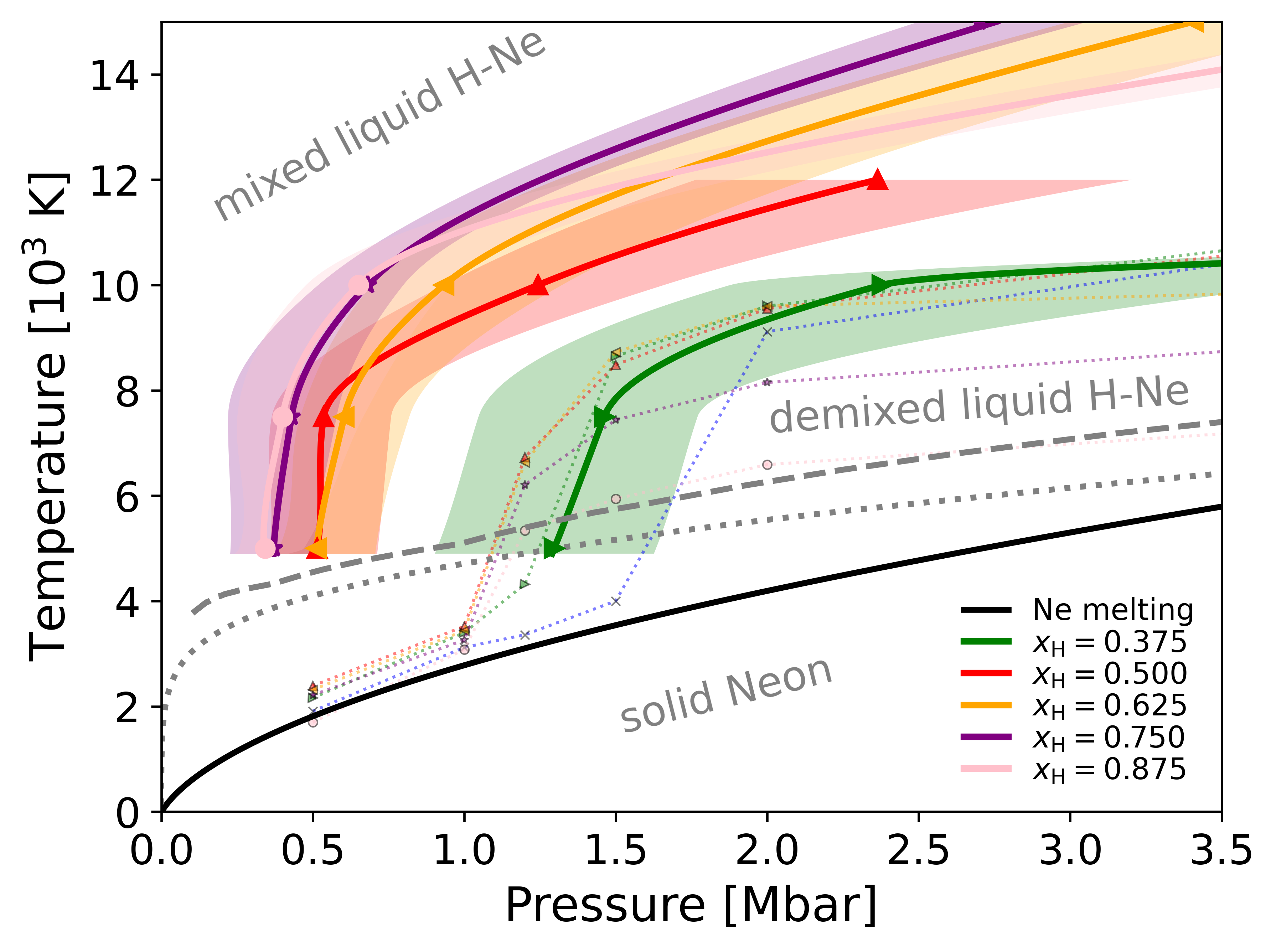}
	\caption{The solid black curve denotes the melting line of Ne. Phase separation in liquid H–Ne occurs at $p$–$T$ conditions below the color-coded curves (corresponding to different H concentrations $x_{\textrm{H}}$). For comparison, phase separation in H–He mixtures takes place below the dotted curves (same color scheme)~\cite{Schoettler2018}. The grey dashed and dotted lines show exemplary isentropes for Jupiter\cite{Militzer2013b} and Saturn~\cite{Nettelmann2013a}, respectively.}
    \label{MiscGap}
\end{figure}

\subsection{Miscibility diagram}\label{sec:MG}
Fig.~\ref{MiscGap} shows the miscibility diagram obtained in this study. Each marker corresponds to a specific $p$–$T$ condition at which the height of the first peak in the H–Ne PDF reaches its maximum. The solid colored lines connect these points for each composition, and phase separation is predicted to occur below these lines, where the color indicates the corresponding H concentration. At temperatures below $10000$~K, phase separation is most pronounced at high H concentrations. As the H concentration $x_\textrm{H}$ decreases, the pressure required to induce phase separation increases. In this regime, Ne promotes the formation of H$_2$ molecules and thereby suppresses H metallization, which is generally associated with the onset of phase separation~(\cite{Lorenzen2011}).

Next, we compare our findings for H--Ne with an earlier prediction for the H--He miscibility gap. The dotted lines in Fig.~\ref{MiscGap} indicate the $p$--$T$ conditions at which H--He mixtures become immiscible, as reported in Ref~(\cite{Schoettler2018}). Overall, phase separation in H--Ne is significantly stronger than in H--He. This difference can be understood in terms of excluded-volume effects and the entropy of mixing. The substantially larger atomic size of Ne introduces stronger packing frustration when combined with H, effectively acting as a large hard sphere that disrupts local H configurations. Studies of binary hard-sphere mixtures have shown that even modest size asymmetry can induce phase separation through positive nonadditivity of the excluded volume, leading to a concave free enthalpy of mixing and ultimately to immiscibility~(\cite{Santos2024}). Interestingly, the composition at which phase separation is strongest also differs between the two systems. In H--He mixtures the largest miscibility gap occurs around $0.5 < x_{\textrm{H}} < 0.625$, whereas in H--Ne it is shifted toward more H-rich mixtures of approx.~$0.750 < x_{\textrm{H}} < 0.875$. This shift is consistent with the strong suppression of H metallization. Because metallic H promotes phase separation, the enhanced stabilization of molecular H$_2$ in the presence of Ne directly modifies the shape and extent of the miscibility gap. Very similar trends are observed in H--He mixtures, although the effect is less pronounced. Despite these differences, both systems exhibit the same qualitative behavior: mixtures with lower H concentrations require substantially higher pressures to undergo phase separation, with the effect being more pronounced in H--Ne.

For reference, Fig.~\ref{MiscGap} also shows representative isentropes of Jupiter~\cite{Militzer2013b} and Saturn~(\cite{Nettelmann2013a}). While H--Ne mixtures are not the primary component of giant planet interiors—where H--He mixtures dominate—the comparison illustrates that the predicted miscibility gap occurs at $p$–$T$ conditions relevant to planetary interiors. Since the physical mechanisms governing phase separation in H--Ne and H--He mixtures are closely related, the H--Ne system serves as a useful surrogate for studying phase separation processes in H-rich planetary interiors.

Although our simulations extend up to $40$~ps, the height of the first H--Ne peak still exhibits statistical fluctuations. To quantify the resulting uncertainty, we compared the pressure at which the first H--Ne peak reaches its maximum with the neighboring lower and higher pressures simulated in this work. These neighboring points provide a natural estimate for the uncertainty in locating the onset of phase separation and define the uncertainty bands shown in Fig.~\ref{MiscGap}. Reducing this uncertainty would require a finer pressure grid, smaller statistical fluctuations, and larger simulation cells, which is computationally demanding within the present first-principles framework. We note that this level of uncertainty is comparable to, or smaller than, what can be resolved in current high-energy-density experiments, where phase separation is typically inferred from macroscopic observables rather than a sharply defined thermodynamic boundary.

We restrict the present study to temperatures of $5000$~K and above, where H--Ne mixtures remain in the warm dense fluid regime and the interplay between molecular dissociation, metallization, transport, and phase separation can be analyzed consistently within the DFT-MD framework. Extending the calculations to lower temperatures would require exploring lower densities to locate the miscibility gap. In plane-wave DFT implementations this rapidly becomes computationally prohibitive, as the larger simulation cells lead to a much smaller Brillouin zone and a correspondingly larger plane-wave basis set. In addition, ionic diffusion—particularly of the heavier Ne atoms—becomes slower at lower temperatures, requiring substantially longer simulation trajectories to obtain statistically converged structural properties.

\begin{figure}
\centering
\includegraphics[width=\hsize]{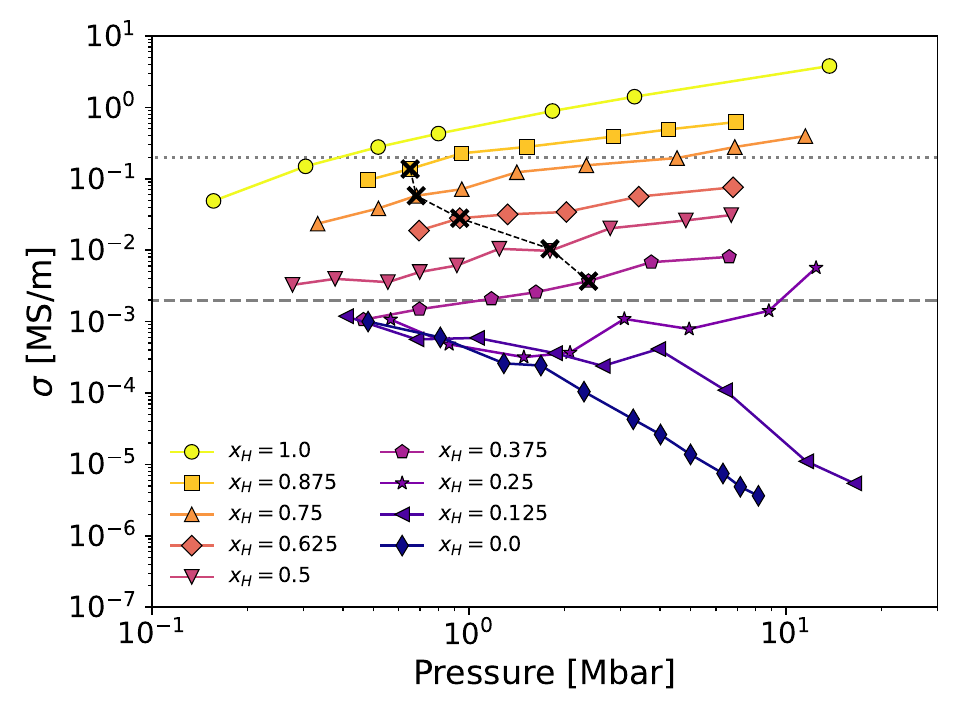}
	\caption{Electrical conductivity of H-Ne mixtures for a temperature of $10000$~K. The $7$ different $x_{\textrm{H}}$ concentrations are color-coded. The gray dashed line depicts the minimum metallic conductivity as derived from the Mott criterion for $T=0$~K, the gray dotted line indicates the corresponding value for fluid H and fluid alkali metals at finite temperatures, and the black dashed line with crosses outlines the pressure where the individual $x_{\textrm{H}}$ concentrations start to phase separate.}
    \label{Fig2}
\end{figure}

\begin{figure}
\centering
\includegraphics[width=\hsize]{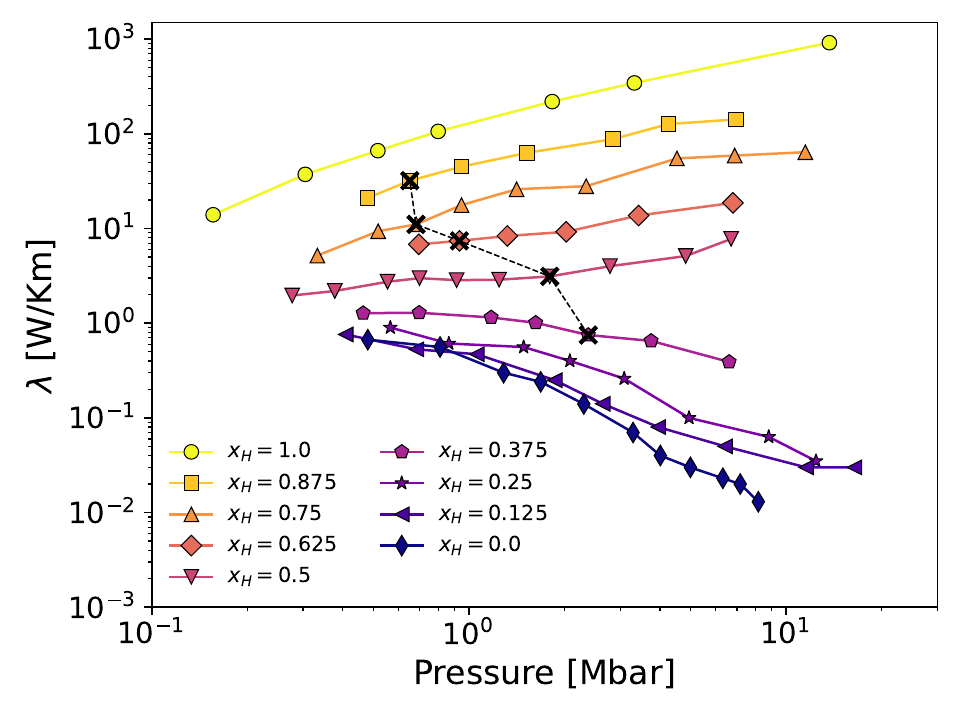}
	\caption{Thermal conductivity of H-Ne mixtures for a temperature of $10000$~K. The $7$ different $x_{\textrm{H}}$ concentrations are color-coded. The black dashed line with crosses outline the pressure where the individual $x_{\textrm{H}}$ concentrations start to phase separate.}
    \label{Fig3}
\end{figure}

\subsection{Electronic Properties}\label{sec:KG}
We calculated the electrical conductivities (Fig.~\ref{Fig2}) using the Kubo-Greenwood formalism~(\cite{French2017, Gajdos2006, Holst2011, Knyazev2013}) to further investigate the behavior of H-Ne mixtures. The zero-temperature Mott criterion is used as a rough indicator for the onset of metallicity. The corresponding minimum metallic conductivity derived from the behavior of doped semiconductors is about $0.002$~MS/m (gray dashed line in Fig.~\ref{Fig2}). Note that the minimum metallic conductivity for fluid H and fluid alkali metals observed in experiments at finite temperatures of few $10^3$~K is higher due to disorder and amounts about $0.2$~MS/m (gray dotted line in Fig.~\ref{Fig2}). For a detailed discussion of the Mott criterion and the metallization transition in various systems, see Ref.~(\cite{Edwards2010}).

As a reference, we find that pure H is metallic according to the zero-temperature Mott criterion for all calculated $p$ (yellow circles), consistent with findings in earlier works~(\cite{Holst2011}). As the concentration of Ne increases, the conductivity decreases significantly, a trend also observed in H–He mixtures~(\cite{Lorenzen2011}). The black dashed line in Fig.~\ref{Fig2} indicates the minimum pressure at which phase separation begins to occur. Therefore, all data points at higher pressures are expected to correspond to phase-separated mixtures, consistent with previous findings showing that the presence of metallic H significantly expands the miscibility gap~(\cite{Lorenzen2011}).

For pressures below $2$~Mbar, the electronic conductivity of H concentrations $x_{\textrm{H}}>0.25$ is equal to pure Ne. Here, H$_2$ remains mostly molecular and insulating and does not lead to an increase in conductivity. In contrast, for pressures above $2$~Mbar, the conductivity of the mixtures exceeds the conductivity of pure Ne. This increase can be attributed to enhanced electronic overlap, polarization effects, and dissociation/ionization of H.

For pure Ne, we find a monotonic decrease in conductivity with increasing pressure. This behavior can be understood by the persistence of the large band gap, which suppresses thermal carrier generation. As a consequence, compression primarily enhances electron–ion scattering rather than increasing the number of free carriers, leading to a reduction in conductivity. The metallization of Ne at finite temperature has not yet been investigated in detail; however, zero-temperature calculations predict gap closure only at pressures on the order of $\sim 2000$~Mbar~(\cite{Tang2017}), i.e., several orders of magnitude higher than the conditions explored in this work.

The trends of the thermal conductivities (Fig.~\ref{Fig3}) closely mirror those observed for the electrical conductivity: Ne strongly suppresses the thermal conductivity in comparison to pure H. Increasing the pressure leads to a further reduction in the thermal conductivity, particularly for high Ne concentrations. Similarly to the electrical conductivity, this behavior can be traced back to the strengthening of H$_2$ bonds in the presence of Ne.

\section{Conclusion}
Mixtures of H, He and/or other noble gases play an important role in modeling the structure and evolution of giant planets such as Jupiter and Saturn. While H–He remains the dominant mixture in planetary interiors, investigating H–Ne mixtures provides valuable insight into H–noble gas mixtures and helps to better understand the physics governing H–He under similar $p$--$T$ conditions. Our simulations reveal that H–Ne mixtures undergo phase separation and show qualitative similarities to H-He mixtures. However, we predict that the minimum pressure required to trigger phase separation is almost an order of magnitude lower than in H–He mixtures.

These findings make H--Ne a powerful experimental surrogate for probing the physics of phase separation in warm dense H-rich mixtures. H-Ne demixing could be directly investigated using small-angle X-ray scattering (SAXS)~(\cite{Glatter1982}), phase contrast imaging~(\cite{Husband2022, Kono2015}), and/or X-ray diffraction (XRD), since the large atomic number of Ne enhances X-ray contrast and reveals density inhomogeneities. Such measurements are considerably more challenging in H--He mixtures~(\cite{Brygoo2021, Loubeyre1987}). Similar experimental techniques have already been successfully employed to study diamond formation in C-H and C-H-O mixtures~(\cite{He2022, Kraus2017}). Additionally, Raman scattering and terahertz spectroscopy could probe the stabilization of H$_2$ molecules induced by the presence of Ne atoms.

The strong phase separation predicted for H-Ne, together with the preferential stabilization of molecular H$_2$ and the suppression of electronic conductivity in the Ne-rich phase, mirrors qualitative trends previously identified in H-He mixtures~(\cite{Karasiev2026, Vorberger2007, Lorenzen2011}). These similarities reinforce the interpretation that phase separation in H-rich mixtures is governed primarily by electronic-structure effects associated with metallic H rather than by simple entropic or mass-based arguments~(\cite{Lorenzen2011, Vorberger2007, Karasiev2026}).

Noble gases, and Ne in particular, are important tracers of formation and evolution processes in giant planets such as Jupiter and Saturn, as they are chemically inert yet sensitive to phase separation in H-rich environments. Although H-Ne mixtures are not directly representative of planetary interiors—where He is the dominant secondary component and Ne is present only as a trace species—our results nonetheless carry important implications for planetary physics~(\cite{Nettelmann2024, Lodders2003, Mandt2020b, Mahaffy2000, Niemann1996, Nettelmann2025}). The strong depletion of Ne observed in Jupiter's atmosphere has been interpreted as evidence that Ne preferentially partitions into helium-rich droplets once H--He phase separation begins~(\cite{Wilson2010}). Our results support this picture by demonstrating that Ne has a strong thermodynamic tendency to separate from H-rich metallic regions under comparable conditions.

More broadly, because direct experimental constraints on H-He miscibility remain limited~(\cite{Brygoo2021, Loubeyre1987}), the experimentally accessible H-Ne system provides an important benchmark for first-principles predictions of phase separation in H-rich mixtures. Experimental validation of the H--Ne miscibility gap and its associated electronic signatures at comparable temperatures but lower pressures would therefore substantially increase confidence in theoretical H-He phase diagrams. H-He phase separation is believed to play a central role in the thermal evolution of gas giant planets—by modifying the radial distribution of composition and entropy, generating stabilizing compositional gradients, and regulating the efficiency of convective heat transport—such validation would directly reduce uncertainties in models of planetary cooling histories, interior stratification, and magnetic field generation~(\cite{Schoettler2018, Bergermann2021a, Morales2013, Lorenzen2009, Mankovich2020, Howard2024, Puestow2016, Fortney2004, Stevenson1975, Guillot1999}).

\begin{acknowledgments}
The authors gratefully acknowledge the computing time made available to them on the high-performance computers Emmy and Lise at the NHR Centers G\"ottingen and Berlin. We thank M.~Knudson, N.~Nettelmann and D.~Kraus for fruitful discussions. RR thanks F.~Hensel and P.~P.~Edwards for many profound discussions on the Mott criterion and the metallization in fluids. Armin Bergermann acknowledges the support of the Alexander von Humboldt Foundation and the Ev. Studienwerk Villigst.
\end{acknowledgments}

\appendix
\twocolumngrid

\section{Pair distribution functions of simulations with $512$ particles}\label{sec:PDFs}
\begin{figure}
\centering
\includegraphics[width=\hsize]{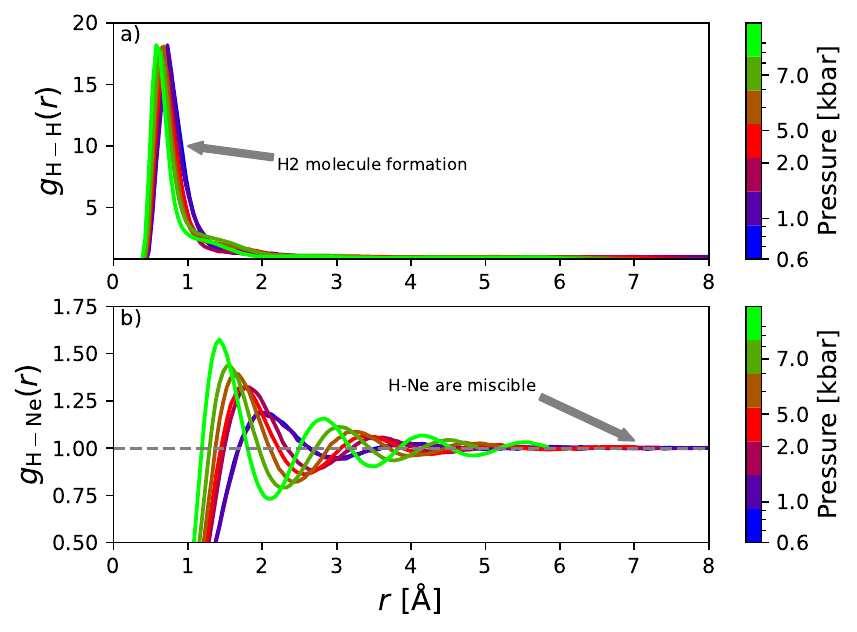}
	\caption{PDFs at a temperature of $T=10000$~K and a H concentration $x_\mathrm{H}=0.125$ with $512$ particles. Panel (a) displays the H–H PDF, while panel (b) shows the H–Ne PDF. Different pressures are distinguished by color coding.}
    \label{fig:Tail16112}
\end{figure}

\begin{figure}
\centering
\includegraphics[width=\hsize]{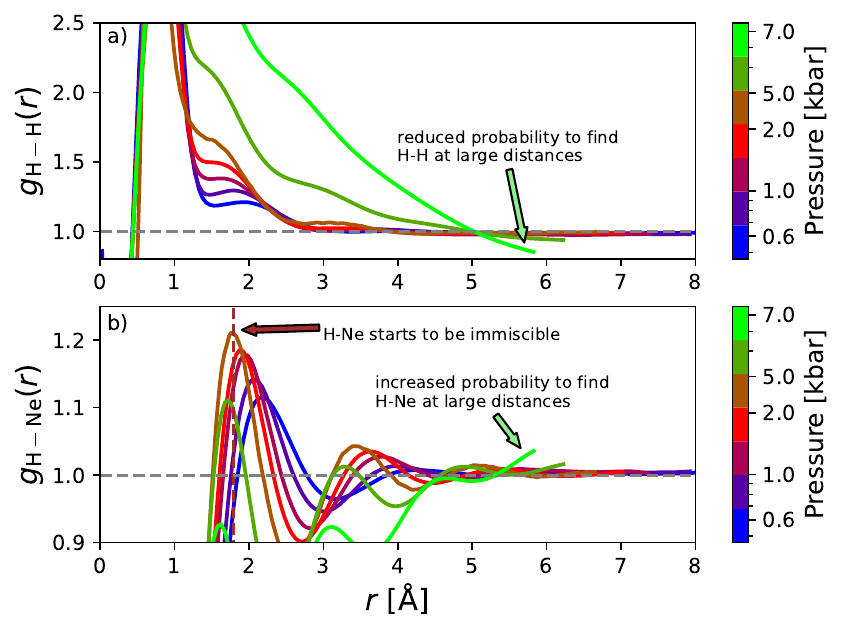}
	\caption{PDFs at a temperature of $T=10000$~K and a H concentration $x_\mathrm{H}=0.375$ with $512$ particles. Panel (a) displays the H–H PDF, while panel (b) shows the H–Ne PDF. Different pressures are distinguished by color coding.}
    \label{fig:Tail4880}
\end{figure}

\begin{figure}
\centering
\includegraphics[width=\hsize]{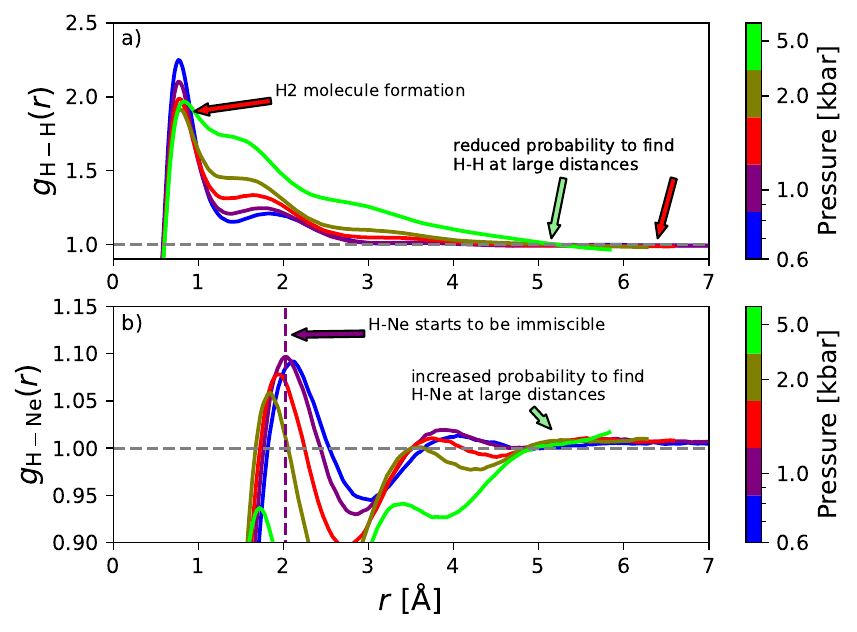}
	\caption{PDFs at a temperature of $10000$~K and a H concentration $x_\mathrm{H}=0.625$ with $512$ particles. Panel (a) displays the H–H PDF, while panel (b) shows the H–Ne PDF. Different pressures are distinguished by color coding.}
    \label{fig:Tail8048}
\end{figure}

\begin{figure}
\centering
\includegraphics[width=\hsize]{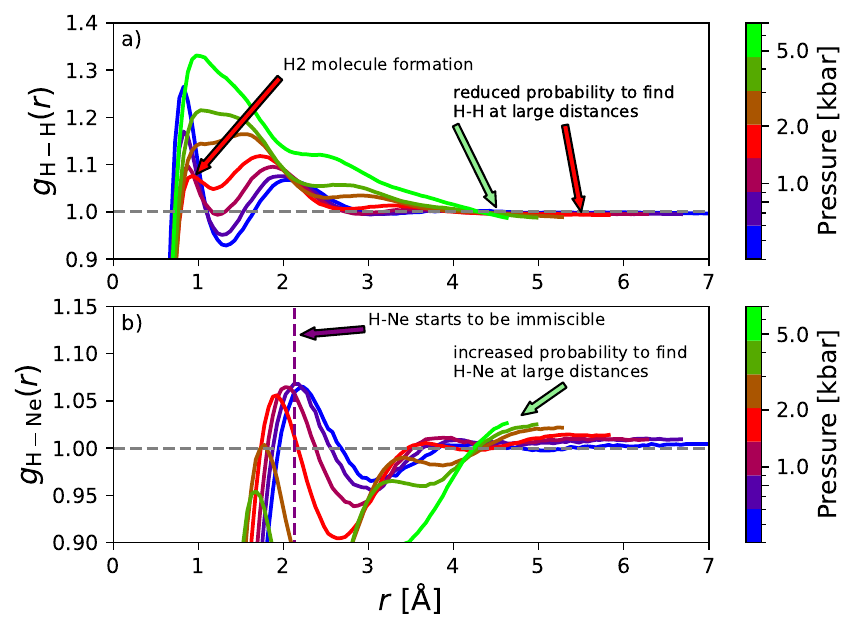}
	\caption{PDFs at a temperature of $10000$~K and a H concentration $x_\mathrm{H}=0.875$ with $512$ particles. Panel (a) displays the H–H PDF, while panel (b) shows the H–Ne PDF. Different pressures are distinguished by color coding.}
    \label{fig:Tail11216}
\end{figure}

We present the H–H and H–Ne PDFs of large simulations with 512 atoms in Figs.~\ref{fig:Tail16112} to~\ref{fig:Tail11216}. We do not include Ne–Ne PDFs in our analysis, since they are less conclusive: the larger effective size of Ne naturally shifts the Ne–Ne peak to larger distances. Note that the long-range behavior of the H--Ne and H--H PDF provides an additional qualitative but reliable indicator for phase separation. In practice, however, its interpretation depends on finite-size effects. Close to the boundary between mixed and demixed states, the interpretation of the PDF tails becomes increasingly ambiguous. For this reason, we use the PDF tail primarily as a consistency check, while the onset of demixing reported in the main text is estimated from the pressure dependence of the first H--Ne peak.

In Fig.~\ref{fig:Tail16112} ($x_{\textrm{H}}=0.125$), we do not find any indications for phase separation. The probabilities of finding H–Ne and H–H pairs both approach 1 at large distances. In addition, the height of the first peak of the H–Ne PDF increases with decreasing pressure. This behavior is consistent with the system remaining fully miscible at all pressures.

In Fig.~\ref{fig:Tail4880}, we show the PDFs for H–H and H–Ne at a H concentration of $x_{\textrm{H}}=0.375$ and find clear evidence of demixing. The first peak of the H–Ne PDF reaches its maximum at $2.378$~Mbar (brown), which we estimate to be pressure where H and Ne begin to lose miscibility. At higher pressures, the tails of the PDFs display pronounced demixing signatures: the probability of finding H–Ne pairs increases markedly above 1 at large separations, while the probability of finding H–H is suppressed (below 1). Note that the heights of the first H–Ne peaks and the long-range tails signal the onset of phase separation at the same pressures. Additionally, the H–H PDF shows signatures of molecule formation at the highest pressures.

In Fig.~\ref{fig:Tail8048}, we present the PDFs for H–H and H–Ne at a H concentration of $x_{\textrm{H}}=0.625$. Again, we find strong indications of phase separation. The first peak of the H–Ne PDF reaches its maximum at $0.933$~Mbar (purple), which we estimate as the pressure where H and Ne begin to lose miscibility. At higher pressures, the tails in the PDFs again display phase separation signatures: the probability of finding H–Ne pairs increases above 1 at large separations, while the probability of finding H–H is suppressed. This effect weakens as the pressure decreases. Comparing the red ($2.025$~Mbar) and purple ($1.321$~Mbar) curves, the interpretation of the tails becomes more ambiguous, making it difficult to pinpoint the precise transition pressure from long-range correlations alone. Closer inspection still reveals subtle hints of phase separation—particularly in the H–H PDF at large distances—that vanish entirely at the lowest pressure. Thus, although the tails are less conclusive, they remain fully consistent with the sharper criterion provided by the H–Ne peak height. 

Finally, we show the PDFs for H–H and H–Ne at a higher H concentration of $x_{\textrm{H}}=0.875$ (Fig.~\ref{fig:Tail11216}). The behavior closely mirrors the case of $x_{\textrm{H}}=0.625$. The main difference is that the H–H peak is less pronounced, reflecting a reduced tendency for the formation of H$_2$ molecules in the presence of fewer Ne atoms. Once again, the height of the first H–Ne peak provides the most robust indicator of phase separation, in excellent agreement with the qualitative trends observed at larger separations.

\section{Height of the first peak in the PDF}\label{sec:conv}
\begin{figure*}[t]
\centering
\includegraphics[scale=0.4]{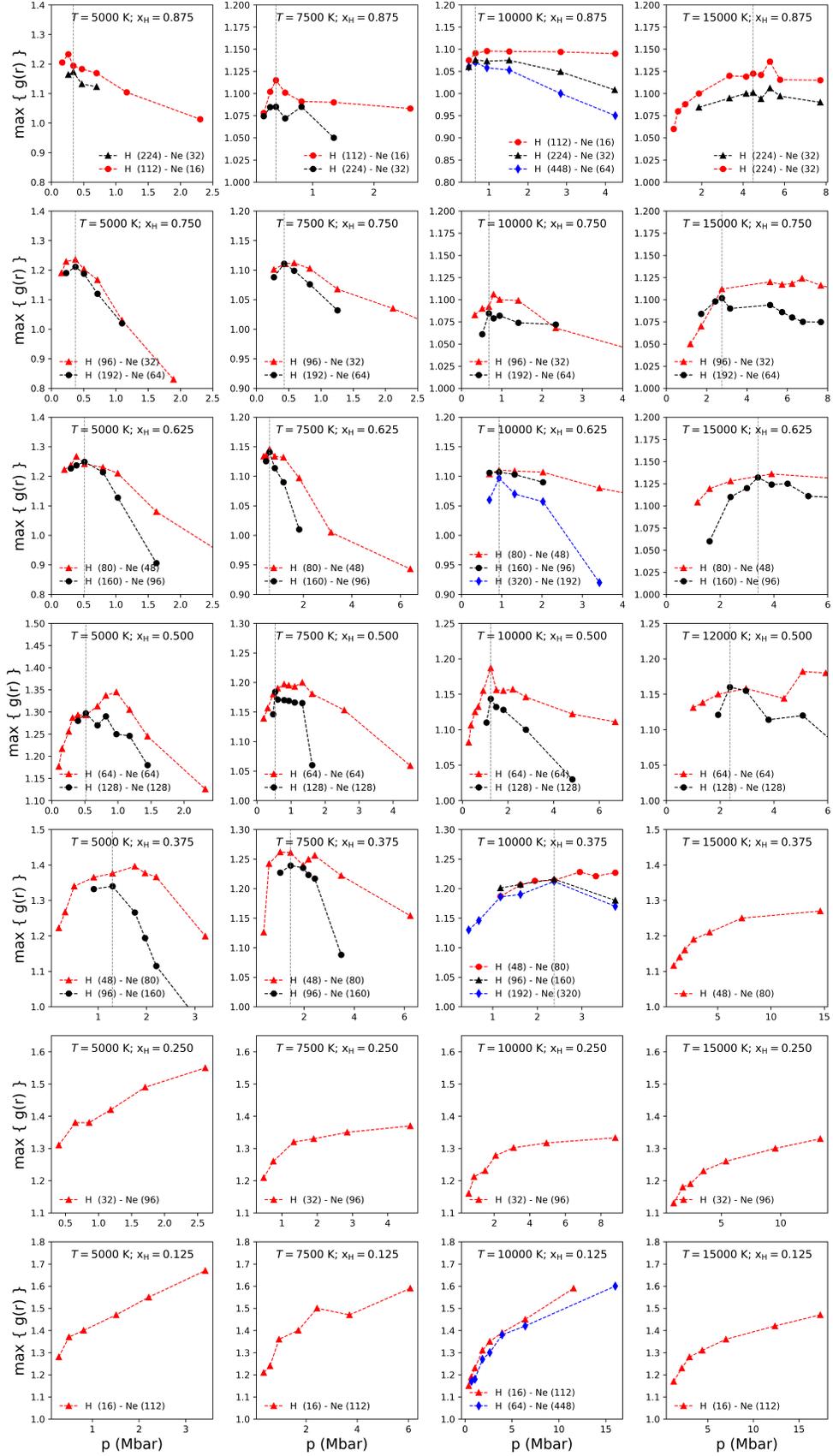}
\caption{Height of the first peak in the PDF for $128$, $256$, and $512$ particles for all $p$-$T$-$x$ conditions investigated in this work. The dashed line shows the pressure corresponding to the estimated onset of phase separation.}
\label{fig:HeightAll}
\end{figure*}

We report the height of the first peak of the H--Ne PDF (Fig.~\ref{fig:HeightAll}) for all the conditions ($p$--$T$--$x_\textrm{H}$) calculated in this work. The signature of phase separation is particularly pronounced at lower temperatures, most notably at $5000$~K, where the variation of the peak height with pressure is significantly more distinct than at higher temperatures. This behavior is expected because these conditions lie deeper within the phase separated region, leading to a clearer structural contrast between the mixed and phase-separated states.

To assess finite-size effects, we performed simulations with $128$ and $256$ particles for all relevant conditions ($p$--$T$--$x_\textrm{H}$) and with $N=512$ particles for selected cases ($T=10000$~K and $x_{\textrm{H}}=0.875, 0.625, 0.375,$ and $0.125$). The uncertainty in locating the onset of phase separation was estimated by comparing the pressure at which the first H--Ne peak reaches its maximum with the neighboring lower and higher pressures simulated in this work. These neighboring points provide a natural estimate of the uncertainty in the phase separation pressure. All results obtained with $128$ atoms fall within this uncertainty range. Nevertheless, to reduce statistical fluctuations and improve the robustness of the structural analysis, we used $256$ atoms for the final results reported in the main text. Note that we did not conduct any further simulations with $256$ particles at H-Ne concentration where we did not find any indications for phase separation given the high computational cost of these simulations.

In some of the smaller simulations ($N=128$ and $N=256$), we observe a second maximum in the peak height as a function of pressure. The maximum at higher pressure corresponds to the onset of H$_2$ dissociation. Karasiev~\textit{et al.}~\cite{Karasiev2026} reported similar behavior for H--He mixtures. This secondary maximum weakens or disappears upon increasing particle number. This can be attributed to the stronger phase separation in larger simulation cells, which reduces the probability of finding either atomic H or molecular H$_2$ in the immediate vicinity of Ne atoms. Finally, all simulation cells were inspected visually to ensure robustness and that the correct peak was identified.

\clearpage
\section{Equation of state}
In the following, we present the EOS results obtained from simulations with $N=128$ particles for all investigated $\rho$-$T$-$x$ conditions. This system size provides reliable EOS data while allowing efficient coverage of the broad range of state points considered.
\begin{table}[h]
\caption{Equation-of-state data for H--Ne mixtures from DFT-MD simulations for $x_H=0.125$. Listed are hydrogen fraction $x_H$, density $\rho$, temperature $T$, pressure $P$, and internal energy per atom $U$.}
\label{tab:eos_hne_xh0125}
\centering
\scriptsize
\setlength{\tabcolsep}{4pt}
\renewcommand{\arraystretch}{1.1}
\begin{tabular}{ccccc}
\toprule
$x_H$ & $\rho$ (g/cm$^3$) & $T$ (K) & $P$ (GPa) & $U$ (eV/atom) \\
\midrule
0.125 & 2.51977 &  5000 &   281.361 &  0.7462679688 \\
0.125 & 3.14971 &  5000 &   499.377 &  1.049210938  \\
0.125 & 3.77965 &  5000 &   813.050 &  1.444875     \\
0.125 & 4.72456 &  5000 &  1509.82  &  2.210664063  \\
0.125 & 5.39950 &  5000 &  2209.83  &  2.900734375  \\
0.125 & 6.29942 &  5000 &  3420.41  &  3.983625     \\\\
\midrule
0.125 & 2.51977 &  7500 &   347.332 &  1.238015625  \\
0.125 & 3.14971 &  7500 &   592.182 &  1.579851563  \\
0.125 & 3.77965 &  7500 &   943.067 &  1.971398438  \\
0.125 & 4.72456 &  7500 &  1695.18  &  2.876904000  \\
0.125 & 5.39950 &  7500 &  2430.84  &  3.530929688  \\
0.125 & 6.29942 &  7500 &  3696.68  &  4.629976563  \\
0.125 & 7.55930 &  7500 &  6075.26  &  6.524226563  \\\\
\midrule
0.125 & 2.51977 & 10000 &   408.782 &  1.749960938  \\
0.125 & 3.14971 & 10000 &   682.801 &  2.089390625  \\
0.125 & 3.77965 & 10000 &  1058.75  &  2.5546875    \\
0.125 & 4.72456 & 10000 &  1864.08  &  3.393953125  \\
0.125 & 5.39950 & 10000 &  2640.38  &  4.153781250  \\
0.125 & 6.29942 & 10000 &  3948.84  &  5.285796875  \\
0.125 & 7.55930 & 10000 &  6414.82  &  7.223648438  \\
0.125 & 9.44913 & 10000 & 11560.8   & 10.67984375   \\\\
\midrule
0.125 & 3.77965 & 15000 &  1273.09  &  3.673359375  \\
0.125 & 4.72456 & 15000 &  2163.25  &  4.577984375  \\
0.125 & 5.39950 & 15000 &  3002.78  &  5.368671875  \\
0.125 & 6.29942 & 15000 &  4408.19  &  6.538382813  \\
0.125 & 7.55930 & 15000 &  7001.81  &  8.553359375  \\
0.125 & 9.44913 & 15000 & 12369.8   & 12.15523438   \\
0.125 & 10.7990 & 15000 & 17343.7   & 15.13351563   \\
\bottomrule
\end{tabular}
\end{table}
\begin{table}[h]
\caption{Same as Table~\ref{tab:eos_hne_xh0125}, but for $x_H=0.25$.}
\label{tab:eos_hne_xh025}
\centering
\scriptsize
\setlength{\tabcolsep}{4pt}
\renewcommand{\arraystretch}{1.1}
\begin{tabular}{ccccc}
\toprule
$x_H$ & $\rho$ (g/cm$^3$) & $T$ (K) & $P$ (GPa) & $U$ (eV/atom) \\
\midrule
0.25 & 2.72505 &  5000 &  403.289 & 0.4733945313 \\
0.25 & 3.63341 &  5000 &  650.114 & 0.7807320313 \\
0.25 & 3.63341 &  5000 &  859.344 & 1.02159375   \\
0.25 & 4.08758 &  5000 & 1184.59  & 1.36503125   \\
0.25 & 4.67152 &  5000 & 1706.01  & 1.87428125   \\
0.25 & 5.45011 &  5000 & 2616.81  & 2.681773438  \\
0.25 & 6.54013 &  5000 & 4311.54  & 4.0186875    \\\\
\midrule
0.25 & 2.72505 &  7500 &  487.157 & 1.031835938  \\
0.25 & 3.27007 &  7500 &  759.186 & 1.34959375   \\
0.25 & 4.08758 &  7500 & 1337.53  & 1.93796875   \\
0.25 & 4.67152 &  7500 & 1903.23  & 2.493054688  \\
0.25 & 5.45011 &  7500 & 2851.04  & 3.320140625  \\
0.25 & 6.54013 &  7500 & 4640.15  & 4.704500000  \\
0.25 & 8.17516 &  7500 & 8390.42  & 7.216937500  \\\\
\midrule
0.25 & 2.72505 & 10000 &  565.111 & 1.596695313  \\
0.25 & 3.27007 & 10000 &  864.336 & 1.92103125   \\
0.25 & 4.08758 & 10000 & 1487.19  & 2.552054688  \\
0.25 & 4.67152 & 10000 & 2079.88  & 3.104562500  \\
0.25 & 5.45011 & 10000 & 3085.55  & 3.982531250  \\
0.25 & 6.54013 & 10000 & 4940.87  & 5.389312500  \\
0.25 & 8.17516 & 10000 & 8813.09  & 7.945078125  \\
0.25 & 9.34305 & 10000 & 12420.1  & 10.06789063  \\\\
\midrule
0.25 & 3.27007 & 15000 & 1060.03  & 3.061679688  \\
0.25 & 4.08758 & 15000 & 1759.32  & 3.740375000  \\
0.25 & 4.67152 & 15000 & 2410.22  & 4.324171875  \\
0.25 & 5.45011 & 15000 & 3507.71  & 5.229570313  \\
0.25 & 6.54013 & 15000 & 5488.62  & 6.723117188  \\
0.25 & 8.17516 & 15000 & 9553.62  & 9.375937500  \\
0.25 & 9.34305 & 15000 & 13331.1  & 11.56625000  \\\\
\bottomrule
\end{tabular}
\end{table}

\begin{table}[h]
\caption{Same as Table~\ref{tab:eos_hne_xh0125}, but for $x_H=0.375$.}
\label{tab:eos_hne_xh0375}
\centering
\scriptsize
\setlength{\tabcolsep}{4pt}
\renewcommand{\arraystretch}{1.1}
\begin{tabular}{ccccc}
\toprule
$x_H$ & $\rho$ (g/cm$^3$) & $T$ (K) & $P$ (GPa) & $U$ (eV/atom) \\
\midrule
0.375 & 1.84032 &  5000 &  187.326 & -0.2275921875 \\
0.375 & 2.30040 &  5000 &  319.839 & -0.040112109  \\
0.375 & 2.76048 &  5000 &  507.066 &  0.1805796875 \\
0.375 & 3.45060 &  5000 &  910.375 &  0.605915625  \\
0.375 & 3.94354 &  5000 & 1297.77  &  0.982953125  \\
0.375 & 4.41677 &  5000 & 1758.03  &  1.39975      \\
0.375 & 4.60080 &  5000 & 1970.31  &  1.57615625   \\
0.375 & 4.80083 &  5000 & 2201.61  &  1.774820313  \\
0.375 & 5.52096 &  5000 & 3215.27  &  2.551328125  \\\\
\midrule
0.375 & 2.30040 &  7500 &  394.124 &  0.5212476563 \\
0.375 & 2.76048 &  7500 &  605.026 &  0.73424375   \\
0.375 & 3.45060 &  7500 & 1047.60  &  1.198648438  \\
0.375 & 3.94354 &  7500 & 1462.94  &  1.595117188  \\
0.375 &	4.24689	&  7500	& 1770.70  &  1.870105469  \\
0.375 & 4.41677 &  7500 & 1964.40  &  2.028101563  \\
0.375 & 4.60080 &  7500 & 2177.74  &  2.216007813  \\
0.375 & 4.80083 &  7500 & 2427.20  &  2.406601562  \\
0.375 & 5.52096 &  7500 & 3483.63  &  3.206453125  \\
0.375 & 6.90120 &  7500 & 6239.43  &  5.029445313  \\\\
\midrule
0.375 & 2.30040 & 10000 &  463.870 &  1.097250000  \\
0.375 & 2.76048 & 10000 &  696.775 &  1.338351563  \\
0.375 & 3.45060 & 10000 & 1175.02  &  1.804257813  \\
0.375 & 3.94354 & 10000 & 1623.21  &  2.207562500  \\
0.375 &	4.24689	& 10000	& 1950.14  &  2.484968750  \\
0.375 & 4.60080 & 10000 & 2378.01  &  2.832000000  \\
0.375 &	5.01905	& 10000	& 2956.11  &  3.281195313  \\
0.375 & 5.25806 & 10000 & 3314.01  &  3.5391875    \\
0.375 & 5.52096 & 10000 & 3751.43  &  3.856773438  \\
0.375 & 5.81154 & 10000 & 4280.10  &  4.223179688  \\
0.275 & 6.49525 & 10000 & 5661.84  &  5.1278125    \\
0.375 & 6.90120 & 10000 & 6610.07  &  5.721718750  \\\\
\midrule
0.375 & 2.76048 & 15000 &  872.339 &  2.471484375  \\
0.375 & 3.45060 & 15000 & 1418.35  &  2.967492188  \\
0.375 & 3.94354 & 15000 & 1916.73  &  3.390820313  \\
0.375 & 4.60080 & 15000 & 2748.14  &  4.055218750  \\
0.375 & 5.52096 & 15000 & 4236.73  &  5.128109375  \\
0.375 & 6.90120 & 15000 & 7266.48  &  7.06934375   \\
0.375 & 9.20160 & 15000 & 14628.3  & 11.0115625    \\
0.375 & 10.0381 & 15000 & 18056.5  & 12.6237500    \\
\bottomrule
\end{tabular}
\end{table}

\begin{table}[h]
\caption{Same as Table~\ref{tab:eos_hne_xh0125}, but for $x_H=0.5$.}
\label{tab:eos_hne_xh05}
\centering
\scriptsize
\setlength{\tabcolsep}{4pt}
\renewcommand{\arraystretch}{1.1}
\begin{tabular}{ccccc}
\toprule
$x_H$ & $\rho$ (g/cm$^3$) & $T$ (K) & $P$ (GPa) & $U$ (eV/atom) \\
\midrule
0.5 & 1.25050 &  5000 &  106.362 & -0.7838515625 \\
0.5 & 1.50060 &  5000 &  156.949 & -0.7122648438 \\
0.5 & 1.87574 &  5000 &  257.913 & -0.5678773438 \\
0.5 & 2.04627 &  5000 &  315.102 & -0.4938960938 \\
0.5 & 2.25089 &  5000 &  395.885 & -0.3860906250 \\
0.5	& 2.36936 &  5000 &  449.857 & -0.315234375	\\
0.5 & 2.50099 &  5000 &  515.806 & -0.2567054688 \\
0.5 & 2.81362 &  5000 &  690.486 & -0.0616116400 \\
0.5 & 3.00119 &  5000 &  818.098 &  0.05333026562 \\
0.5 & 3.21556 &  5000 &  976.817 &  0.2111546875 \\
0.5 & 3.46291 &  5000 & 1175.75  &  0.3977718750 \\
0.5 & 3.75149 &  5000 & 1448.28  &  0.6400062500 \\
0.5 & 4.50179 &  5000 & 2321.35  &  1.333976563  \\\\
\midrule
0.5 & 1.50060 &  7500 &  200.400 & -0.1429203125 \\
0.5 & 1.87574 &  7500 &  317.795 &  0.0219589840 \\
0.5 & 2.25089 &  7500 &  477.453 &  0.2059984380 \\
0.5	& 2.36936 &  7500 &  536.289 &  0.262573828	 \\
0.5 & 2.50099 &  7500 &  608.597 &  0.3436796880 \\
0.5 & 2.81362 &  7500 &  799.714 &  0.5360820310 \\
0.5 & 3.00119 &  7500 &  936.991 &  0.6614906250 \\
0.5 & 3.21556 &  7500 & 1109.12  &  0.8218437500 \\
0.5 & 3.46291 &  7500 & 1331.70  &  1.012500000  \\
0.5 & 3.75149 &  7500 & 1623.64  &  1.251031250  \\
0.5 & 4.50179 &  7500 & 2569.09  &  1.955656250  \\
0.5 & 5.62723 &  7500 & 4507.74  &  3.205445313  \\\\
\midrule
0.5 & 1.60778 & 10000 &  277.183 &  0.4881843750 \\
0.5 & 1.87574 & 10000 &  377.609 &  0.6090757813 \\
0.5 & 2.25089 & 10000 &  553.927 &  0.7822343750 \\
0.5 & 2.50099 & 10000 &  698.180 &  0.9234453130 \\
0.5 & 2.81362 & 10000 &  914.377 &  1.129257813  \\
0.5 & 3.21556 & 10000 & 1244.63  &  1.407085938  \\
0.5 & 3.46291 & 10000 & 1485.64  &  1.607148438  \\
0.5 & 3.75149 & 10000 & 1798.16  &  1.850515625  \\
0.5 & 4.09253 & 10000 & 2211.13  &  2.152468750  \\
0.5 & 4.50179 & 10000 & 2783.96  &  2.552390625  \\
0.5 & 5.62723 & 10000 & 4824.17  &  3.840109375  \\
0.5 & 6.43112 & 10000 & 6707.44  &  4.915328125  \\\\
\midrule
0.5	& 2.81362 & 12000 &  999.73  & 1.574421875	\\
0.5	& 3.21556 & 12000 & 1352.25  & 1.87121875	\\
0.5	& 3.75149 & 12000 & 1928.2   & 2.314632813	\\
0.5	& 4.09253 & 12000 & 2363.76	 & 2.631355469  \\
0.5 & 4.50179 & 12000 & 2964.95	 & 3.045851563	\\
0.5	& 5.00199 & 12000 & 3802.92	 & 3.578320313	\\
0.5 & 5.29622 & 12000 & 4376.64  & 3.948953125	\\
0.5 & 5.62723 & 12000 & 5071.95	 & 4.375539063	\\
0.5 & 6.00238 & 12000 & 5928.86  & 4.867601563  \\
0.5	& 7.50298 & 12000 & 10166.3	 & 7.049359375	\\\\
\bottomrule
\end{tabular}
\end{table}

\begin{table}[h]
\caption{Same as Table~\ref{tab:eos_hne_xh0125}, but for $x_H=0.625$.}
\label{tab:eos_hne_xh0625}
\centering
\scriptsize
\setlength{\tabcolsep}{4pt}
\renewcommand{\arraystretch}{1.1}
\begin{tabular}{ccccc}
\toprule
$x_H$ & $\rho$ (g/cm$^3$) & $T$ (K) & $P$ (GPa) & $U$ (eV/atom) \\
\midrule
0.625 & 1.45109 &  5000 &  200.087 & -1.043445310 \\
0.625 & 1.74131 &  5000 &  301.132 & -0.908632813 \\
0.625 & 1.93479 &  5000 &  387.578 & -0.81365625  \\
0.625 & 2.17663 &  5000 &  512.975 & -0.676190625 \\
0.625 & 2.48758 &  5000 &  708.913 & -0.468804688 \\
0.625 & 2.90218 &  5000 & 1038.67  & -0.217172656 \\
0.625 & 3.48261 &  5000 & 1627.10  &  0.3050757813 \\
0.625 & 4.35327 &  5000 & 2857.11  &  1.1137890600 \\\\
\midrule
0.625 & 1.74131 &  7500 &  367.541 & -0.293510156 \\
0.625 & 1.93479 &  7500 &  458.771 & -0.2002976563 \\
0.625 & 2.17663 &  7500 &  604.345 &  0.06736984375 \\
0.625 & 2.48758 &  7500 &  818.578 &  0.1236726563 \\
0.625 & 2.90218 &  7500 & 1181.40  &  0.4194507813 \\
0.625 & 3.48261 &  7500 & 1827.60  &  0.8902578125 \\
0.625 & 4.35327 &  7500 & 3141.68  &  1.702867188 \\
0.625 & 5.80436 &  7500 & 6406.36  &  3.425585940 \\\\
\midrule
0.625 & 2.17663 & 10000 &  695.003 &  0.5016804688 \\
0.625 & 2.48758 & 10000 &  933.798 &  0.6906054690 \\
0.625 & 2.90218 & 10000 & 1321.60  &  0.9825546860 \\
0.625 & 3.48261 & 10000 & 2025.66  &  1.458164063 \\
0.625 & 4.35327 & 10000 & 3425.65  &  2.310929688 \\
0.625 & 5.80436 & 10000 & 6796.00  &  4.046468750 \\\\
\midrule
0.625 & 2.48758 & 15000 & 1159.27  &  1.764570313 \\
0.625 & 2.90218 & 15000 & 1606.62  &  2.076390625 \\
0.625 & 3.48261 & 15000 & 2387.26  &  2.585523438 \\
0.625 &	3.86957	& 15000	& 2996.35  &  2.887234375 \\
0.625 &	4.09719	& 15000	& 3408.23  &  3.133945313 \\
0.625 & 4.35327 & 15000 & 3916.34  &  3.473414063 \\
0.625 &	4.64349	& 15000	& 4508.77  &  3.743882813 \\
0.625 &	4.97516	& 15000	& 5279.87  &  4.152031250 \\
0.625 & 5.52796 & 15000 & 6727.71  &  4.904898438 \\
0.625 & 5.80436 & 15000 & 7545.32  &  5.320976563 \\
0.625 & 6.10985 & 15000 & 8474.33  &  5.737398438 \\
0.625 & 6.96523 & 15000 & 11429.7  &  6.991695313 \\
\bottomrule
\end{tabular}
\end{table}

\begin{table}[h]
\caption{Same as Table~\ref{tab:eos_hne_xh0125}, but for $x_H=0.75$.}
\label{tab:eos_hne_xh075}
\centering
\scriptsize
\setlength{\tabcolsep}{4pt}
\renewcommand{\arraystretch}{1.1}
\begin{tabular}{ccccc}
\toprule
$x_H$ & $\rho$ (g/cm$^3$) & $T$ (K) & $P$ (GPa) & $U$ (eV/atom) \\
\midrule
0.75 & 1.02643 &  5000 &  155.336 & -1.4973046900 \\
0.75 & 1.23172 &  5000 &  228.304 & -1.4083671880 \\
0.75 & 1.53965 &  5000 &  371.604 & -1.2228437500 \\
0.75 & 1.75960 &  5000 &  503.328 & -1.0806875000 \\
0.75 & 2.05287 &  5000 &  714.636 & -0.8694140625 \\
0.75 & 2.46344 &  5000 & 1094.42  & -0.5568465630 \\
0.75 & 3.07930 &  5000 & 1893.58  & -0.03757523438 \\\\
\midrule
0.75 & 1.23172 &  7500 &  277.100 & -0.774654688  \\
0.75 & 1.53965 &  7500 &  437.400 & -0.6142742188 \\
0.75 & 1.75960 &  7500 &  585.832 & -0.4890773438 \\
0.75 & 2.05287 &  7500 &  827.415 & -0.3036328125 \\
0.75 & 2.46344 &  7500 & 1257.21  & -0.0106212500 \\
0.75 & 3.07930 &  7500 & 2119.99  &  0.5054101563 \\
0.75 & 4.10574 &  7500 & 4191.87  &  1.55321875   \\\\
\midrule
0.75 & 1.23172 & 10000 &  332.656 & -0.2085203125 \\
0.75 & 1.53965 & 10000 &  517.268 & -0.0722185160 \\
0.75 & 1.75960 & 10000 &  681.887 &  0.04477457813 \\
0.75 & 1.89496 & 10000 &  798.019 &  0.1209234375 \\
0.75 & 2.05287 & 10000 &  947.450 &  0.2223500000 \\
0.75 & 2.46344 & 10000 & 1413.35  &  0.5116031250 \\
0.75 & 3.07930 & 10000 & 2341.55  &  1.043804687  \\
0.75 & 4.10574 & 10000 & 4524.4	  &  2.120046875  \\
0.75 & 6.15861 & 10000 & 11489.1  &  4.857960937  \\\\
\midrule
0.75 & 1.53965 & 15000 &  681.278 & 0.940062500  \\
0.75 & 1.75960 & 15000 &  878.473 & 1.054652500  \\
0.75 & 2.05287 & 15000 & 1188.91  & 1.230396063  \\
0.75 & 2.46344 & 15000 & 1725.08  & 1.541078125  \\
0.75 & 3.07930 & 15000 & 2757.98  & 2.098210938  \\
0.75 & 4.10574 & 15000 & 5142.58  & 3.245773438  \\
0.75 & 4.32183 & 15000 & 5742.44  & 3.499312500  \\
0.75 & 4.47898 & 15000 & 6237.28  & 3.710932148  \\ 
0.75 & 4.64800 & 15000 & 6733.01  & 3.919894531  \\
0.75 & 4.92688 & 15000 & 7662.85  & 4.308742188  \\
0.75 & 6.15861 & 15000 & 12494.3  & 6.069523438  \\\\
\bottomrule
\end{tabular}
\end{table}

\begin{table}[h]
\caption{Same as Table~\ref{tab:eos_hne_xh0125}, but for $x_H=0.875$.}
\label{tab:eos_hne_xh0875}
\centering
\scriptsize
\setlength{\tabcolsep}{4pt}
\renewcommand{\arraystretch}{1.1}
\begin{tabular}{ccccc}
\toprule
$x_H$ & $\rho$ (g/cm$^3$) & $T$ (K) & $P$ (GPa) & $U$ (eV/atom) \\
\midrule
0.875 & 0.722135 &  5000 &  166.114 & -1.877085937 \\
0.875 & 0.902670 &  5000 &  261.001 & -1.74253125  \\
0.875 & 1.031620 &  5000 &  339.408 & -1.631125000 \\
0.875 & 1.203560 &  5000 &  471.209 & -1.484453125 \\
0.875 & 1.444270 &  5000 &  700.395 & -1.283351563 \\
0.875 & 1.805340 &  5000 & 1167.64  & -0.97571875  \\
0.875 & 2.407120 &  5000 & 2306.01  & -0.948135185 \\\\
\midrule
0.875 & 0.722135 &  7500 &  201.848 & -2.912611111 \\
0.875 & 0.902669 &  7500 &  306.117 & -1.119132813 \\
0.875 & 1.031620 &  7500 &  400.950 & -1.045757813 \\
0.875 & 1.203560 &  7500 &  549.560 & -0.943828125 \\
0.875 & 1.444270 &  7500 &  816.921 & -0.7734125   \\
0.875 & 1.805340 &  7500 & 1344.88  & -0.4829570313 \\
0.875 & 2.407120 &  7500 & 2592.30  &  0.1106164063 \\\\
\midrule
0.875 & 1.031620 & 10000 &  478.604 & -0.5487304688 \\
0.875 & 1.203560 & 10000 &  651.355 & -0.4538632813 \\
0.875 & 1.444270 & 10000 &  946.649 & -0.29999757813 \\
0.875 & 1.805340 & 10000 & 1522.90  & -0.00899484    \\
0.875 & 2.407120 & 10000 & 2849.06  &  0.593537500    \\
0.875 & 2.888540 & 10000 & 4256.43  &  1.174221094    \\
0.875 & 3.610680 & 10000 & 6945.73  &  2.143117188    \\\\
\midrule
0.875 & 1.031620 & 15000 &  651.730 & 0.398275      \\
0.875 & 1.203560 & 15000 &  859.486 & 0.477223437   \\
0.875 & 1.444270 & 15000 & 1210.35  & 0.628070313   \\
0.875 & 1.805340 & 15000 & 1868.94  & 0.923984375   \\
0.875 & 2.407120 & 15000 & 3345.64  & 1.555765625   \\
0.875 & 2.674570 & 15000 & 4151.15  & 1.875968750   \\
0.875 & 2.777440 & 15000 & 4485.14  & 2.002000000   \\
0.875 & 2.888540 & 15000 & 4865.35  & 2.14709375    \\
0.875 & 3.008900 & 15000 & 5297.96  & 2.309296875   \\
0.875 & 3.139720 & 15000 & 5788.37  & 2.484171875   \\
0.875 & 3.610680 & 15000 & 7729.75  & 3.145859375   \\
\bottomrule
\end{tabular}
\end{table}

\clearpage\newpage
\vspace{100cm}

\bibliography{statphys}

@article{Karasiev2026,
      title={Inhibiting Conduction by He Mixing in Interiors of Jupiter and Saturn}, 
      author={Valentin V. Karasiev and S. X. Hu and Joshua P. Hinz and R. M. N. Goshadze and Shuai Zhang and Armin Bergermann and Ronald Redmer},
      year={2026},
      eprint={2601.23152},
      archivePrefix={arXiv},
      primaryClass={astro-ph.EP},
      url={https://arxiv.org/abs/2601.23152}, 
}

@article{Nettelmann2025,
doi = {10.3847/PSJ/adbdb7},
url = {https://doi.org/10.3847/PSJ/adbdb7},
year = {2025},
month = {apr},
publisher = {The American Astronomical Society},
volume = {6},
number = {4},
pages = {98},
author = {Nettelmann, N. and Fortney, J. J.},
title = {Jupiter’s Interior with an Inverted Helium Gradient},
journal = {The Planetary Science Journal}
}

@article{Bonitz2024,
    author = {Bonitz, Michael and Vorberger, Jan and Bethkenhagen, Mandy and Böhme, Maximilian P. and Ceperley, David M. and Filinov, Alexey and Gawne, Thomas and Graziani, Frank and Gregori, Gianluca and Hamann, Paul and Hansen, Stephanie B. and Holzmann, Markus and Hu, S. X. and Kählert, Hanno and Karasiev, Valentin V. and Kleinschmidt, Uwe and Kordts, Linda and Makait, Christopher and Militzer, Burkhard and Moldabekov, Zhandos A. and Pierleoni, Carlo and Preising, Martin and Ramakrishna, Kushal and Redmer, Ronald and Schwalbe, Sebastian and Svensson, Pontus and Dornheim, Tobias},
    title = {Toward first principles-based simulations of dense hydrogen},
    journal = {Physics of Plasmas},
    volume = {31},
    number = {11},
    pages = {110501},
    year = {2024},
    month = {11},
    issn = {1070-664X},
    doi = {10.1063/5.0219405},
    url = {https://doi.org/10.1063/5.0219405}
}

@article{Bund2021,
  title = {Isotope Quantum Effects in the Metallization Transition in Liquid Hydrogen},
  author = {van de Bund, Sebastiaan and Wiebe, Heather and Ackland, Graeme J.},
  journal = {Phys. Rev. Lett.},
  volume = {126},
  issue = {22},
  pages = {225701},
  numpages = {6},
  year = {2021},
  month = {Jun},
  publisher = {American Physical Society},
  doi = {10.1103/PhysRevLett.126.225701},
  url = {https://link.aps.org/doi/10.1103/PhysRevLett.126.225701}
}

@Article{Chang2024,
author={Chang, Xiaoju
and Chen, Bo
and Zeng, Qiyu
and Wang, Han
and Chen, Kaiguo
and Tong, Qunchao
and Yu, Xiaoxiang
and Kang, Dongdong
and Zhang, Shen
and Guo, Fangyu
and Hou, Yong
and Zhao, Zengxiu
and Yao, Yansun
and Ma, Yanming
and Dai, Jiayu},
title={Theoretical evidence of H-He demixing under Jupiter and Saturn conditions},
journal={Nature Communications},
year={2024},
month={Oct},
day={02},
volume={15},
number={1},
pages={8543},
issn={2041-1723},
doi={10.1038/s41467-024-52868-4},
url={https://doi.org/10.1038/s41467-024-52868-4}
}

@article{Geng2019,
  title = {Thermodynamic anomalies and three distinct liquid-liquid transitions in warm dense liquid hydrogen},
  author = {Geng, Hua Y. and Wu, Q. and Marqu\'es, Miriam and Ackland, Graeme J.},
  journal = {Phys. Rev. B},
  volume = {100},
  issue = {13},
  pages = {134109},
  numpages = {11},
  year = {2019},
  month = {Oct},
  publisher = {American Physical Society},
  doi = {10.1103/PhysRevB.100.134109},
  url = {https://link.aps.org/doi/10.1103/PhysRevB.100.134109}
}

@article{Mazzola2018,
  title = {Phase Diagram of Hydrogen and a Hydrogen-Helium Mixture at Planetary Conditions by Quantum Monte Carlo Simulations},
  author = {Mazzola, Guglielmo and Helled, Ravit and Sorella, Sandro},
  journal = {Phys. Rev. Lett.},
  volume = {120},
  issue = {2},
  pages = {025701},
  numpages = {6},
  year = {2018},
  month = {Jan},
  publisher = {American Physical Society},
  doi = {10.1103/PhysRevLett.120.025701},
  url = {https://link.aps.org/doi/10.1103/PhysRevLett.120.025701}
}

@article{Howard2024,
	author = {{Howard, S.} and {Müller, S.} and {Helled, R.}},
	title = {Evolution of Jupiter and Saturn with helium rain},
	DOI= "10.1051/0004-6361/202450629",
	url= "https://doi.org/10.1051/0004-6361/202450629",
	journal = {Astron. Astrophys.},
	year = 2024,
	volume = 689,
	pages = "A15",
}

@Article{Mandt2020b,
author={Mandt, K. E.
and Mousis, O.
and Lunine, J.
and Marty, B.
and Smith, T.
and Luspay-Kuti, A.
and Aguichine, A.},
title={Tracing the Origins of the Ice Giants Through Noble Gas Isotopic Composition},
journal={Space Science Reviews},
year={2020},
month={Jul},
day={23},
volume={216},
number={5},
pages={99},
issn={1572-9672},
doi={10.1007/s11214-020-00723-5},
url={https://doi.org/10.1007/s11214-020-00723-5}
}

@article{Mahaffy2000,
author = {Mahaffy, P. R. and Niemann, H. B. and Alpert, A. and Atreya, S. K. and Demick, J. and Donahue, T. M. and Harpold, D. N. and Owen, T. C.},
title = {Noble gas abundance and isotope ratios in the atmosphere of Jupiter from the Galileo Probe Mass Spectrometer},
journal = {Journal of Geophysical Research: Planets},
volume = {105},
number = {E6},
pages = {15061-15071},
doi = {https://doi.org/10.1029/1999JE001224},
url = {https://agupubs.onlinelibrary.wiley.com/doi/abs/10.1029/1999JE001224},
eprint = {https://agupubs.onlinelibrary.wiley.com/doi/pdf/10.1029/1999JE001224},
year = {2000}
}

@article{Knudson2015,
  title = {Direct observation of an abrupt insulator-to-metal transition in dense liquid deuterium},
  author = {M. D. Knudson and M. P. Desjarlais and A. Becker and R. W. Lemke and K. R. Cochrane and M. E. Savage and D. E. Bliss and T. R. Mattsson and R. Redmer},
  journal = {Science},
  volume = {348},
  issue = {6242},
  pages = {1455--1460},
  year = {2015},
  month = {June},
  doi = {10.1126/science.aaa7471},
  url = {https://science.sciencemag.org/content/348/6242/1455}
}

@article{Pierleoni2016,
  title = {Liquid--liquid phase transition in hydrogen by coupled electron--ion Monte Carlo simulations},
  author = {Carlo Pierleoni and Miguel A. Morales and Giovanni Rillo and David M. Ceperley},
  journal = {Proceedings of the National Academy of Sciences},
  volume = {113},
  issue = {18},
  pages = {4953--4957},
  year = {2016},
  month = {April},
  doi = {10.1073/pnas.1603853113}
}

@article{Hinz2020,
  title = {Fully consistent density functional theory determination of the insulator-metal transition boundary in warm dense hydrogen},
  author = {Joshua Hinz and Valentin V. Karasiev and S. X. Hu and Mohamed Zaghoo and Daniel Mejía-Rodríguez and S. B. Trickey and L. Calderín},
  journal = {Phys. Rev. Res.},
  volume = {2},
  pages = {032065(R)},
  year = {2020},
  month = {September},
  doi = {10.1103/PhysRevResearch.2.032065},
  received = {7 February 2020},
  revised = {11 May 2020},
  accepted = {20 August 2020},
  published = {10 September 2020}
}

@article{Celliers2018,
  title = {Insulator-metal transition in dense fluid deuterium},
  author = {Peter M. Celliers and Marius Millot and Stephanie Brygoo and R. Stewart McWilliams and Dayne E. Fratanduono and J. Ryan Rygg and Alexander F. Goncharov and Paul Loubeyre and Jon H. Eggert and J. Luc Peterson and Nathan B. Meezan and Sebastien Le Pape and Gilbert W. Collins and Raymond Jeanloz and Russell J. Hemley},
  journal = {Science},
  volume = {361},
  issue = {6403},
  pages = {677--682},
  year = {2018},
  month = {August},
  doi = {10.1126/science.aat0970},
  pmid = {30115805},
  url = {https://www.sciencemag.org/content/361/6403/677}
}

@Article{Helled2020b,
  author={Helled, Ravit and Mazzola, Guglielmo and Redmer, Ronald},
  title={Understanding dense hydrogen at planetary conditions},
  journal={Nature Reviews Physics},
  year={2020},
  month={Oct},
  day={01},
  volume={2},
  number={10},
  pages={562-574},
  issn={2522-5820},
  doi={10.1038/s42254-020-0223-3},
  url={https://doi.org/10.1038/s42254-020-0223-3}
}

@article{Cheng2023,
  author = {Bingqing Cheng and Sebastien Hamel and Mandy Bethkenhagen},
  title = {Thermodynamics of diamond formation from hydrocarbon mixtures in planets},
  journal = {Nature Commun.},
  volume = {14},
  number = {1},
  pages = {1104},
  year = {2023},
  doi = {10.1038/s41467-023-36841-1}
}

@book{Oliveira2017,
  title={Equilibrium Thermodynamics},
  author={Mario J. de Oliveira},
  isbn={978-3-662-53205-8},
  series={Graduate Texts in Physics},
  year={2017},
  publisher={Springer},
  url = {https://link.springer.com/book/10.1007/978-3-662-53207-2}
}

@article{Bergermann2024,
  title = {Ab initio calculation of the miscibility diagram for mixtures of hydrogen and water},
  author = {Bergermann, Armin and French, Martin and Redmer, Ronald},
  journal = {Phys. Rev. B},
  volume = {109},
  issue = {17},
  pages = {174107},
  numpages = {6},
  year = {2024},
  month = {May},
  publisher = {American Physical Society},
  doi = {10.1103/PhysRevB.109.174107},
  url = {https://link.aps.org/doi/10.1103/PhysRevB.109.174107}
}

@article{Mankovich2020,
	year = {2020},
	volume = {889},
	pages = {51},
	author = {Christopher R. Mankovich and Jonathan J. Fortney},
	title = {Evidence for a Dichotomy in the Interior Structures of Jupiter and Saturn from Helium Phase Separation},
	journal = {Astrophys. J.},
	doi = {10.3847/1538-4357/ab6210}
}

@article{Knyazev2013,
  title = {Ab initio calculation of transport and optical properties of aluminum: Influence of simulation parameters},
  journal = {Comp. Mater. Sci.},
  volume = {79},
  pages = {817-829},
  year = {2013},
  issn = {0927-0256},
  doi = {https://doi.org/10.1016/j.commatsci.2013.04.066},
  url = {https://www.sciencedirect.com/science/article/pii/S0927025613002541},
  author = {D.V. Knyazev and P.R. Levashov}
}

@Article{Bergermann2021b,
  author ={Bergermann, Armin and French, Martin and Redmer, Ronald},
  title  ={Gibbs-ensemble Monte Carlo simulation of H2–H2O mixtures},
  journal  ={Phys. Chem. Chem. Phys.},
  year  ={2021},
  volume  ={23},
  issue  ={22},
  pages  ={12637-12643},
  publisher  ={The Royal Society of Chemistry},
  doi  ={10.1039/D1CP00515D},
  url  ={http://dx.doi.org/10.1039/D1CP00515D},
}

@article{Bergermann2021a,
  title = {Gibbs-ensemble Monte Carlo simulation of H2-He mixtures},
  author = {Bergermann, Armin and French, Martin and Sch\"ottler, Manuel and Redmer, Ronald},
  journal = {Phys. Rev. E},
  volume = {103},
  issue = {1},
  pages = {013307},
  numpages = {6},
  year = {2021},
  month = {Jan},
  publisher = {American Physical Society},
  doi = {10.1103/PhysRevE.103.013307},
  url = {https://link.aps.org/doi/10.1103/PhysRevE.103.013307}
}

@article{He2022,
  author = {Zhiyu He  and Melanie Rödel  and Julian Lütgert  and Armin Bergermann  and Mandy Bethkenhagen  and Deniza Chekrygina  and Thomas E. Cowan  and Adrien Descamps  and Martin French  and Eric Galtier  and Arianna E. Gleason  and Griffin D. Glenn  and Siegfried H. Glenzer  and Yuichi Inubushi  and Nicholas J. Hartley  and Jean-Alexis Hernandez  and Benjamin Heuser  and Oliver S. Humphries  and Nobuki Kamimura  and Kento Katagiri  and Dimitri Khaghani  and Hae Ja Lee  and Emma E. McBride  and Kohei Miyanishi  and Bob Nagler  and Benjamin Ofori-Okai  and Norimasa Ozaki  and Silvia Pandolfi  and Chongbing Qu  and Divyanshu Ranjan  and Ronald Redmer  and Christopher Schoenwaelder  and Anja K. Schuster  and Michael G. Stevenson  and Keiichi Sueda  and Tadashi Togashi  and Tommaso Vinci  and Katja Voigt  and Jan Vorberger  and Makina Yabashi  and Toshinori Yabuuchi  and Lisa M. V. Zinta  and Alessandra Ravasio  and Dominik Kraus },
  title = {Diamond formation kinetics in shock-compressed C-H-O samples recorded by small-angle x-ray scattering and x-ray diffraction},
  journal = {Sci. Adv.},
  volume = {8},
  number = {35},
  pages = {eabo0617},
  year = {2022},
  doi = {10.1126/sciadv.abo0617},
  URL = {https://www.science.org/doi/abs/10.1126/sciadv.abo0617},
}

@article{Kraus2017,
  author = {D. Kraus and J. Vorberger and A. Pak and N. J. Hartley and L. B. Fletcher and S. Frydrych and E. Galtier and E. J. Gamboa and D. O. Gericke and S. H. Glenzer and E. Granados and M. J. MacDonald and A. J. MacKinnon and E. E. McBride and I. Nam and P. Neumayer and M. Roth and A. M. Saunders and A. K. Schuster and P. Sun and T. van Driel and T. Döppner and R. W. Falcone},
  title = {Formation of diamonds in laser-compressed hydrocarbons at planetary interior conditions},
  journal = {Nature Astronomy},
  volume = {1},
  number = {9},
  pages = {606-611},
  year = {2017},
  doi = {10.1038/s41550-017-0219-9},
}

@article{Militzer2021,
  title = {First-principles equation of state database for warm dense matter computation},
  author = {Militzer, Burkhard and Gonz\'alez-Cataldo, Felipe and Zhang, Shuai and Driver, Kevin P. and Soubiran, Fran\ifmmode \mbox{\c{c}}\else \c{c}\fi{}ois},
  journal = {Phys. Rev. E},
  volume = {103},
  issue = {1},
  pages = {013203},
  numpages = {13},
  year = {2021},
  month = {Jan},
  publisher = {American Physical Society},
  doi = {10.1103/PhysRevE.103.013203},
  url = {https://link.aps.org/doi/10.1103/PhysRevE.103.013203}
}

@article{Driver2015,
  title = {First-principles simulations and shock Hugoniot calculations of warm dense neon},
  author = {Driver, K. P. and Militzer, B.},
  journal = {Phys. Rev. B},
  volume = {91},
  issue = {4},
  pages = {045103},
  numpages = {8},
  year = {2015},
  month = {Jan},
  publisher = {American Physical Society},
  doi = {10.1103/PhysRevB.91.045103},
  url = {https://link.aps.org/doi/10.1103/PhysRevB.91.045103}
}

@article{Kono2015,
    author = {Kono, Yoshio and Kenney-Benson, Curtis and Shibazaki, Yuki and Park, Changyong and Wang, Yanbin and Shen, Guoyin},
    title = {X-ray imaging for studying behavior of liquids at high pressures and high temperatures using Paris-Edinburgh press},
    journal = {Review of Scientific Instruments},
    volume = {86},
    number = {7},
    pages = {072207},
    year = {2015},
    month = {07},
    issn = {0034-6748},
    doi = {10.1063/1.4927227},
    url = {https://doi.org/10.1063/1.4927227},
    eprint = {https://pubs.aip.org/aip/rsi/article-pdf/doi/10.1063/1.4927227/15945661/072207\_1\_online.pdf},
}

@article{Husband2022,
    author = {Husband, R. J. and Hagemann, J. and O’Bannon, E. F. and Liermann, H.-P. and Glazyrin, K. and Sneed, D. T. and Lipp, M. J. and Schropp, A. and Evans, W. J. and Jenei, Zs.},
    title = {Simultaneous imaging and diffraction in the dynamic diamond anvil cell},
    journal = {Review of Scientific Instruments},
    volume = {93},
    number = {5},
    pages = {053903},
    year = {2022},
    month = {05},
    issn = {0034-6748},
    doi = {10.1063/5.0084480},
    url = {https://doi.org/10.1063/5.0084480},
    eprint = {https://pubs.aip.org/aip/rsi/article-pdf/doi/10.1063/5.0084480/16633753/053903\_1\_online.pdf},
}

@book{Glatter1982,
  title={Small Angle X-ray Scattering},
  author={Glatter, O. and Kratky, O.},
  url={https://books.google.com/books?id=6TNsxgEACAAJ},
  year={1982},
  publisher={Blackwell Science}
}

@article{Santos2024,
  title = {Demixing can occur in binary hard-sphere mixtures with negative nonadditivity},
  author = {Santos, A. and L\'opez de Haro, M.},
  journal = {Phys. Rev. E},
  volume = {72},
  issue = {1},
  pages = {010501},
  numpages = {4},
  year = {2005},
  month = {Jul},
  publisher = {American Physical Society},
  doi = {10.1103/PhysRevE.72.010501},
  url = {https://link.aps.org/doi/10.1103/PhysRevE.72.010501}
}

@article{Bergermann2024b,
    author = {Bergermann, Armin and Kleindienst, Lucas and Redmer, Ronald},
    title = {Nonmetal-to-metal transition in liquid hydrogen using density functional theory and the Heyd–Scuseria–Ernzerhof exchange-correlation functional},
    journal = {The Journal of Chemical Physics},
    volume = {161},
    number = {23},
    pages = {234303},
    year = {2024},
    month = {12},
    issn = {0021-9606},
    doi = {10.1063/5.0241111},
    url = {https://doi.org/10.1063/5.0241111},
    eprint = {https://pubs.aip.org/aip/jcp/article-pdf/doi/10.1063/5.0241111/20306452/234303\_1\_5.0241111.pdf},
}

@article{Arridge2014,
title = {The science case for an orbital mission to Uranus: Exploring the origins and evolution of ice giant planets},
journal = {Planetary and Space Science},
volume = {104},
pages = {122-140},
year = {2014},
note = {Surfaces, atmospheres and magnetospheres of the outer planets and their satellites and ring systems: Part X},
issn = {0032-0633},
doi = {https://doi.org/10.1016/j.pss.2014.08.009},
url = {https://www.sciencedirect.com/science/article/pii/S0032063314002335},
author = {C.S. Arridge and N. Achilleos and J. Agarwal and C.B. Agnor and R. Ambrosi and N. André and S.V. Badman and K. Baines and D. Banfield and M. Barthélémy and M.M. Bisi and J. Blum and T. Bocanegra-Bahamon and B. Bonfond and C. Bracken and P. Brandt and C. Briand and C. Briois and S. Brooks and J. Castillo-Rogez and T. Cavalié and B. Christophe and A.J. Coates and G. Collinson and J.F. Cooper and M. Costa-Sitja and R. Courtin and I.A. Daglis and I. {de Pater} and M. Desai and D. Dirkx and M.K. Dougherty and R.W. Ebert and G. Filacchione and L.N. Fletcher and J. Fortney and I. Gerth and D. Grassi and D. Grodent and E. Grün and J. Gustin and M. Hedman and R. Helled and P. Henri and S. Hess and J.K. Hillier and M.H. Hofstadter and R. Holme and M. Horanyi and G. Hospodarsky and S. Hsu and P. Irwin and C.M. Jackman and O. Karatekin and S. Kempf and E. Khalisi and K. Konstantinidis and H. Krüger and W.S. Kurth and C. Labrianidis and V. Lainey and L.L. Lamy and M. Laneuville and D. Lucchesi and A. Luntzer and J. MacArthur and A. Maier and A. Masters and S. McKenna-Lawlor and H. Melin and A. Milillo and G. Moragas-Klostermeyer and A. Morschhauser and J.I. Moses and O. Mousis and N. Nettelmann and F.M. Neubauer and T. Nordheim and B. Noyelles and G.S. Orton and M. Owens and R. Peron and C. Plainaki and F. Postberg and N. Rambaux and K. Retherford and S. Reynaud and E. Roussos and C.T. Russell and A.M. Rymer and R. Sallantin and A. Sánchez-Lavega and O. Santolik and J. Saur and K.M. Sayanagi and P. Schenk and J. Schubert and N. Sergis and E.C. Sittler and A. Smith and F. Spahn and R. Srama and T. Stallard and V. Sterken and Z. Sternovsky and M. Tiscareno and G. Tobie and F. Tosi and M. Trieloff and D. Turrini and E.P. Turtle and S. Vinatier and R. Wilson and P. Zarka}
}

@article{Masters2014,
title = {Neptune and Triton: Essential pieces of the Solar System puzzle},
journal = {Planetary and Space Science},
volume = {104},
pages = {108-121},
year = {2014},
note = {Surfaces, atmospheres and magnetospheres of the outer planets and their satellites and ring systems: Part X},
issn = {0032-0633},
doi = {https://doi.org/10.1016/j.pss.2014.05.008},
url = {https://www.sciencedirect.com/science/article/pii/S0032063314001354},
author = {A. Masters and N. Achilleos and C.B. Agnor and S. Campagnola and S. Charnoz and B. Christophe and A.J. Coates and L.N. Fletcher and G.H. Jones and L. Lamy and F. Marzari and N. Nettelmann and J. Ruiz and R. Ambrosi and N. Andre and A. Bhardwaj and J.J. Fortney and C.J. Hansen and R. Helled and G. Moragas-Klostermeyer and G. Orton and L. Ray and S. Reynaud and N. Sergis and R. Srama and M. Volwerk}
}

@article{Gupta2025,
doi = {10.3847/2041-8213/adb631},
url = {https://dx.doi.org/10.3847/2041-8213/adb631},
year = {2025},
month = {mar},
publisher = {The American Astronomical Society},
volume = {982},
number = {2},
pages = {L35},
author = {Gupta, Akash and Stixrude, Lars and Schlichting, Hilke E.},
title = {The Miscibility of Hydrogen and Water in Planetary Atmospheres and Interiors},
journal = {The Astrophysical Journal Letters}
}

@Article{Brygoo2021,
  author={Brygoo, S. and Loubeyre, P. and Millot, M. and Rygg, J. R. and Celliers, P. M. and Eggert, J. H. and Jeanloz, R. and Collins, G. W.},
  title={Evidence of hydrogen helium immiscibility at Jupiter interior conditions},
  journal={Nature},
  year={2021},
  month={May},
  day={01},
  volume={593},
  number={7860},
  pages={517-521},
  issn={1476-4687},
}

@Article{Edwards2010,
  Title                    = {{`{\ldots} a metal conducts and a non-metal doesn't'}},
  Author                   = {P. P. Edwards and M. T. J. Lodge and F. Hensel and R. Redmer},
  Journal                  = {Phil. Trans. R. Soc. A},
  Year                     = {2010},
  Pages                    = {941},
  Volume                   = {368},
  Doi                      = {10.1098/rsta.2009.0282}
}

@Article{Fortney2004,
  Title                    = {Effects of helium phase separation on the evolution of extrasolar giant planets},
  Author                   = {Jonathan J. Fortney and William B. Hubbard},
  Journal                  = {Astrophys. J.},
  Year                     = {2004},
  Number                   = {2, Part 1},
  Pages                    = {1039-1049},
  Volume                   = {608},
  Address                  = {{1427 E 60TH ST, CHICAGO, IL 60637-2954 USA}},
  Affiliation              = {{Fortney, JJ (Reprint Author), Univ Arizona, Lunar \& Planetary Lab, Tucson, AZ 85721 USA. Univ Arizona, Lunar \& Planetary Lab, Tucson, AZ 85721 USA.}},
  Doi                      = {10.1086/420765},
  Journal-iso              = {{Astrophys. J.}},
  Publisher                = {{UNIV CHICAGO PRESS}}
}

@Article{French2017,
  author  = {French,Martin and Redmer,Ronald},
  title   = {Electronic transport in partially ionized water plasmas},
  journal = {Phys. Plasmas},
  year    = {2017},
  volume  = {24},
  number  = {9},
  pages   = {092306},
  doi     = {10.1063/1.4998753},
}

@Article{Gajdos2006,
  Title                    = {Linear optical properties in the projector-augmented wave methodology},
  Author                   = {Gajdo\ifmmode \check{s}\else \v{s}\fi{}, M. and Hummer, K. and Kresse, G. and Furthm\"uller, J. and Bechstedt, F.},
  Journal                  = {Phys. Rev. B},
  Year                     = {2006},

  Month                    = {Jan},
  Pages                    = {045112},
  Volume                   = {73},

  Doi                      = {10.1103/PhysRevB.73.045112},
  Issue                    = {4},
  Numpages                 = {9},
  Publisher                = {American Physical Society},
  Url                      = {https://link.aps.org/doi/10.1103/PhysRevB.73.045112}
}

@Article{Rauer2014,
author={Rauer, H.
and Catala, C.
and Aerts, C.
and Appourchaux, T.
and Benz, W.
and Brandeker, A.
and Christensen-Dalsgaard, J.
and Deleuil, M.
and Gizon, L.
and Goupil, M.-J.
and G{\"u}del, M.
and Janot-Pacheco, E.
and Mas-Hesse, M.
and Pagano, I.
and Piotto, G.
and Pollacco, D.
and Santos
and Smith, A.
and Su{\'a}rez, J.-C.
and Szab{\'o}, R.
and Udry, S.
and Adibekyan, V.
and Alibert, Y.
and Almenara, J.-M.
and Amaro-Seoane, P.
and Eiff, M. Ammler-von
and Asplund, M.
and Antonello, E.
and Barnes, S.
and Baudin, F.
and Belkacem, K.
and Bergemann, M.
and Bihain, G.
and Birch, A. C.
and Bonfils, X.
and Boisse, I.
and Bonomo, A. S.
and Borsa, F.
and Brand{\~a}o, I. M.
and Brocato, E.
and Brun, S.
and Burleigh, M.
and Burston, R.
and Cabrera, J.
and Cassisi, S.
and Chaplin, W.
and Charpinet, S.
and Chiappini, C.
and Church, R. P.
and Csizmadia, Sz.
and Cunha, M.
and Damasso, M.
and Davies, M. B.
and Deeg, H. J.
and D{\'i}az, R. F.
and Dreizler, S.
and Dreyer, C.
and Eggenberger, P.
and Ehrenreich, D.
and Eigm{\"u}ller, P.
and Erikson, A.
and Farmer, R.
and Feltzing, S.
and Oliveira Fialho, F. de
and Figueira, P.
and Forveille, T.
and Fridlund, M.
and Garc{\'i}a, R. A.
and Giommi, P.
and Giuffrida, G.
and Godolt, M.
and da Silva, J. Gomes
and Granzer, T.
and Grenfell, J. L.
and Grotsch-Noels, A.
and G{\"u}nther, E.
and Haswell, C. A.
and Hatzes, A. P.
and H{\'e}brard, G.
and Hekker, S.
and Helled, R.
and Heng, K.
and Jenkins, J. M.
and Johansen, A.
and Khodachenko, M. L.
and Kislyakova, K. G.
and Kley, W.
and Kolb, U.
and Krivova, N.
and Kupka, F.
and Lammer, H.
and Lanza, A. F.
and Lebreton, Y.
and Magrin, D.
and Marcos-Arenal, P.
and Marrese, P. M.
and Marques, J. P.
and Martins, J.
and Mathis, S.
and Mathur, S.
and Messina, S.
and Miglio, A.
and Montalban, J.
and Montalto, M.
and P. F. G. Monteiro, M. J.
and Moradi, H.
and Moravveji, E.
and Mordasini, C.
and Morel, T.
and Mortier, A.
and Nascimbeni, V.
and Nelson, R. P.
and Nielsen, M. B.
and Noack, L.
and Norton, A. J.
and Ofir, A.
and Oshagh, M.
and Ouazzani, R.-M.
and P{\'a}pics, P.
and Parro, V. C.
and Petit, P.
and Plez, B.
and Poretti, E.
and Quirrenbach, A.
and Ragazzoni, R.
and Raimondo, G.
and Rainer, M.
and Reese, D. R.
and Redmer, R.
and Reffert, S.
and Rojas-Ayala, B.
and Roxburgh, I. W.
and Salmon, S.
and Santerne, A.
and Schneider, J.
and Schou, J.
and Schuh, S.
and Schunker, H.
and Silva-Valio, A.
and Silvotti, R.
and Skillen, I.
and Snellen, I.
and Sohl, F.
and Sousa, S. G.
and Sozzetti, A.
and Stello, D.
and Strassmeier, K. G.
and {\v{S}}vanda, M.
and Szab{\'o}, Gy. M.
and Tkachenko, A.
and Valencia, D.
and Van Grootel, V.
and Vauclair, S. D.
and Ventura, P.
and Wagner, F. W.
and Walton, N. A.
and Weingrill, J.
and Werner, S. C.
and Wheatley, P. J.
and Zwintz, K.},
title={The PLATO 2.0 mission},
journal={Experimental Astronomy},
year={2014},
month={Nov},
day={01},
volume={38},
number={1},
pages={249-330},
issn={1572-9508},
doi={10.1007/s10686-014-9383-4},
url={https://doi.org/10.1007/s10686-014-9383-4}
}

@article{Militzer2024,
author = {Burkhard Militzer },
title = {Phase separation of planetary ices explains nondipolar magnetic fields of Uranus and Neptune},
journal = {Proceedings of the National Academy of Sciences},
volume = {121},
number = {49},
pages = {e2403981121},
year = {2024},
doi = {10.1073/pnas.2403981121},
URL = {https://www.pnas.org/doi/abs/10.1073/pnas.2403981121},
eprint = {https://www.pnas.org/doi/pdf/10.1073/pnas.2403981121}}

@article{Cano2024,
	author = {{Cano Amoros, M.} and {Nettelmann, N.} and {Tosi, N.} and {Baumeister, P.} and {Rauer, H.}},
	title = {H2–H2O demixing in Uranus and Neptune: Adiabatic structure models},
	DOI= "10.1051/0004-6361/202452148",
	url= "https://doi.org/10.1051/0004-6361/202452148",
	journal = {Astron. Astrophys.},
	year = 2024,
	volume = 692,
	pages = "A152",
}

@book{Syrkin1964,
  author       = {Syrkin, Y. K. and Dyatkina, M. E.},
  title        = {Structure of Molecules and the Chemical Bond},
  translator   = {Partridge, M. A. and Jordan, D. O.},
  publisher    = {Dover Publications},
  address      = {New York},
  year         = {1964},
  pages        = {ix + 509},
  isbn         = {978-0486611679},
  note         = {Translated and revised from the Russian edition}
}

@article{Howard2025,
      title={The Possibility of Hydrogen-Water Demixing in Uranus, Neptune, K2-18b and TOI-270d}, 
      author={Saburo Howard and Ravit Helled and Armin Bergermann and Ronald Redmer},
      year={2025},
      journal={Astron. Astrophys.},
      volume = {703},
      number = {A154},
      primaryClass={astro-ph.EP},
}

@article{Pin2023,
    author = {Chew, Pin Yu and Reinhardt, Aleks},
    title = {Phase diagrams—Why they matter and how to predict them},
    journal = {The Journal of Chemical Physics},
    volume = {158},
    number = {3},
    pages = {030902},
    year = {2023},
    month = {01},
    issn = {0021-9606},
    doi = {10.1063/5.0131028},
}

@Article{Nettelmann2024,
author={Nettelmann, Nadine
and Cano Amoros, Marina
and Tosi, Nicola
and Helled, Ravit
and Fortney, Jonathan J.},
title={Atmospheric Helium Abundances in the Giant Planets},
journal={Space Science Reviews},
year={2024},
month={Jul},
day={29},
volume={220},
number={5},
pages={56},
issn={1572-9672},
doi={10.1007/s11214-024-01090-1},
url={https://doi.org/10.1007/s11214-024-01090-1}
}

@Article{Guillot1999,
  Title                    = {Interiors of giant planets inside end outside the solar system},
  Author                   = {Tristan Guillot},
  Journal                  = {Science},
  Year                     = {1999},
  Pages                    = {72},
  Volume                   = {286},
  Doi                      = {10.1126/science.286.5437.72}
}

@Article{Holst2011,
  Title                    = {Electronic transport coefficients from ab initio simulations and application to dense liquid hydrogen},
  Author                   = {Holst, Bastian and French, Martin and Redmer, Ronald},
  Journal                  = {Phys. Rev. B},
  Year                     = {2011},
  Pages                    = {235120},
  Volume                   = {83},
  Doi                      = {10.1103/PhysRevB.83.235120}
}

@Article{Marley1995,
  Title                    = {Monte Carlo interior models for Uranus and Neptune},
  Author                   = {Marley, Mark S. and Gómez, Percy and Podolak, Morris},
  Journal                  = {J. Geophys. Res.},
  Year                     = {1995},

  Month                    = nov,
  Number                   = {E11},
  Pages                    = {23349--23353},
  Volume                   = {100},
  Comment                  = {https://doi.org/10.1029/95JE02362},
  ISSN                     = {0148-0227},
  Publisher                = {John Wiley \& Sons, Ltd},
  Url                      = {https://doi.org/10.1029/95JE02362}
}

@article{Kirkwood1935,
  author 		= {Kirkwood,John G. },
  title 		= {Statistical Mechanics of Fluid Mixtures},
  journal 		= {J. Chem. Phys.},
  volume 		= {3},
  number 		= {5},
  pages 		= {300-313},
  year 			= {1935},
  doi 			= {10.1063/1.1749657},
}

@Article{Knudson2001,
  Title                    = {Equation of State Measurements in Liquid Deuterium to 70 GPa},
  Author                   = {M.~D.~Knudson and D.~L.~Hanson and J.~E.~Bailey and C.~A.~Hall and J.~R.~Asay and W.~W.~Anderson},
  Journal                  = {Phys. Rev. Lett.},
  Year                     = {2001},
  Pages                    = {225501},
  Volume                   = {87},
  Doi                      = {10.1103/PhysRevLett.87.225501}
}

@Article{Koci2007a,
  Title                    = {Ab initio and classical molecular dynamics of neon melting at high pressure},
  Author                   = {Ko\ifmmode \check{c}\else \v{c}\fi{}i, L. and Ahuja, R. and Belonoshko, A. B.},
  Journal                  = {Phys. Rev. B},
  Year                     = {2007},

  Month                    = {Jun},
  Pages                    = {214108},
  Volume                   = {75},

  Doi                      = {10.1103/PhysRevB.75.214108},
  Issue                    = {21},
  Numpages                 = {7},
  Publisher                = {American Physical Society},
  Url                      = {https://link.aps.org/doi/10.1103/PhysRevB.75.214108}
}

@Article{Kresse1996a,
  Title                    = {Efficient iterative schemes for ab initio total-energy calculations using a plane-wave basis set},
  Author                   = {G.~Kresse and J.~Furthm\"uller},
  Journal                  = {Phys. Rev. B},
  Year                     = {1996},
  Pages                    = {11169},
  Volume                   = {54},
  Doi                      = {10.1103/PhysRevB.54.11169}
}

@Article{Kresse1996b,
  Title                    = {Efficiency of ab-initio total energy calculations for metals and semiconductors using a plane-wave basis set},
  Author                   = {G. Kresse and J. Furthmüller},
  Journal                  = {Comp. Mater. Sci.},
  Year                     = {1996},
  Number                   = {1},
  Pages                    = {15 - 50},
  Volume                   = {6},

  Doi                      = {https://doi.org/10.1016/0927-0256(96)00008-0},
  ISSN                     = {0927-0256},
  Url                      = {http://www.sciencedirect.com/science/article/pii/0927025696000080}
}

@Article{Kresse1994,
  Title                    = {Ab initio molecular-dynamics simulation of the liquid-metal--amorphous-semiconductor transition in germanium},
  Author                   = {G.~Kresse and J.~Hafner},
  Journal                  = {Phys. Rev. B},
  Year                     = {1994},
  Pages                    = {14251},
  Volume                   = {49},
  Doi                      = {10.1103/PhysRevB.49.14251}
}

@Article{Kresse1993,
  Title                    = {Ab initio molecular dynamics for liquid metals},
  Author                   = {G.~Kresse and J.~Hafner},
  Journal                  = {Phys. Rev. B},
  Year                     = {1993},
  Pages                    = {558},
  Volume                   = {47},
  Doi                      = {10.1103/PhysRevB.47.558}
}

@Article{Lodders2003,
  Title                    = {Solar System Abundances and Condensation Temperatures of the Elements},
  Author                   = {Katharina Lodders},
  Journal                  = {Astrophys. J.},
  Year                     = {2003},
  Pages                    = {1220},
  Volume                   = {591},
  Doi                      = {10.1086/375492}
}

@Article{Lorenzen2011,
  Title                    = {Metallization in hydrogen-helium mixtures},
  Author                   = {Winfried Lorenzen and Bastian Holst and Ronald Redmer},
  Journal                  = {Phys. Rev. B},
  Year                     = {2011},
  Pages                    = {235109},
  Volume                   = {84},
  Doi                      = {10.1103/PhysRevB.84.235109}
}

@Article{Lorenzen2010,
  Title                    = {First-order liquid-liquid phase transition in dense hydrogen},
  Author                   = {Winfried Lorenzen and Bastian Holst and Ronald Redmer},
  Journal                  = {Phys. Rev. B},
  Year                     = {2010},
  Pages                    = {195107},
  Volume                   = {82},
  Doi                      = {10.1103/PhysRevB.82.195107}
}

@Article{Lorenzen2009,
  Title                    = {Demixing of Hydrogen and Helium at Megabar Pressures},
  Author                   = {Lorenzen, Winfried and Holst, Bastian and Redmer, Ronald},
  Journal                  = {Phys. Rev. Lett.},
  Year                     = {2009},
  Pages                    = {115701},
  Volume                   = {102},
  Doi                      = {10.1103/PhysRevLett.102.115701}
}

@Article{Loubeyre1987,
  Title                    = {Binary phase diagram of H$_2$-He mixtures at high temperature and high pressure},
  Author                   = {P. Loubeyre and R. {Le~Toullec} and J. P. Pinceaux},
  Journal                  = {Phys. Rev. B},
  Year                     = {1987},
  Pages                    = {3723},
  Volume                   = {36},
  Doi                      = {10.1103/PhysRevB.36.3723}
}

@Article{McWilliams2015,
  author    = {McWilliams, R. Stewart and Dalton, D. Allen and Kon{\^o}pkov{\'a}, Zuzana and Mahmood, Mohammad F. and Goncharov, Alexander F.},
  title     = {Opacity and conductivity measurements in noble gases at conditions of planetary and stellar interiors},
  journal   = {Proc. Natl. Acad. Sci. USA},
  year      = {2015},
  volume    = {112},
  number    = {26},
  pages     = {7925--7930},
  issn      = {0027-8424},
  doi       = {10.1073/pnas.1421801112},
  publisher = {National Academy of Sciences},
  url       = {https://www.pnas.org/content/112/26/7925},
}

@Article{Militzer2013b,
  Title                    = {Ab Initio Equation of State for Hydrogen-Helium Mixtures with Recalibration of the Giant-Planet Mass-Radius Relation},
  Author                   = {Militzer, B. and Hubbard, W.B.},
  Journal                  = {Astrophys. J.},
  Year                     = {2013},
  Pages                    = {148},
  Volume                   = {774}
}

@Article{Monkhorst1976,
  Title                    = {Special points for Brillouin-zone integrations},
  Author                   = {Hendrik J. Monkhorst and James D. Pack},
  Journal                  = {Phys. Rev. B},
  Year                     = {1976},
  Pages                    = {5188},
  Volume                   = {13},
  Doi                      = {10.1103/PhysRevB.13.5188}
}

@Article{Morales2013,
  Title                    = {Nuclear Quantum Effects and Nonlocal Exchange-Correlation Functionals Applied to Liquid Hydrogen at High Pressure},
  Author                   = {Morales, Miguel A. and McMahon, Jeffrey M. and Pierleoni, Carlo and Ceperley, D. M.},
  Journal                  = {Phys. Rev. Lett.},
  Year                     = {2013},
  Pages                    = {065702},
  Volume                   = {110},
  Doi                      = {10.1103/PhysRevLett.110.065702}
}

@Article{Morales2010a,
  Title                    = {Evidence for a first-order liquid-liquid transition in high-pressure hydrogen from ab initio simulations},
  Author                   = {Morales, Miguel A. and Pierleoni, Carlo and Schwegler, Eric and Ceperley, D. M.},
  Journal                  = {Proc. Natl. Acad. Sci. USA},
  Year                     = {2010},
  Pages                    = {12799-12803},
  Volume                   = {107},
  Doi                      = {10.1073/pnas.1007309107}
}

@Article{Morales2009,
  Title                    = {Phase separation in hydrogen -- helium mixtures at {M}bar pressures},
  Author                   = {Morales, Miguel A. and Schwegler, Eric and Ceperley, David and Pierleoni, Carlo and Hamel, Sebastien and Caspersen, Kyle},
  Journal                  = {Proc. Natl. Acad. Sci. USA},
  Year                     = {2009},
  Number                   = {5},
  Pages                    = {1324-1329},
  Volume                   = {106},
  Doi                      = {10.1073/pnas.0812581106}
}

@article{Morales2013b,
  title = {Hydrogen-helium demixing from first principles: From diamond anvil cells to planetary interiors},
  author = {Morales, Miguel A. and Hamel, Sebastien and Caspersen, Kyle and Schwegler, Eric},
  journal = {Phys. Rev. B},
  volume = {87},
  issue = {17},
  pages = {174105},
  numpages = {4},
  year = {2013},
  month = {May},
  publisher = {American Physical Society},
  doi = {10.1103/PhysRevB.87.174105},
  url = {https://link.aps.org/doi/10.1103/PhysRevB.87.174105}
}

@Article{Nettelmann2013a,
  Title                    = {Saturn layered structure and homogeneous evolution models with different EOS},
  Author                   = {N. Nettelmann and R. P\"ustow and R. Redmer},
  Journal                  = {Icarus},
  Year                     = {2013},
  Pages = {548-557},
  Volume = {225}
}

@Article{Perdew1996,
  Title                    = {Generalized Gradient Approximation Made Simple},
  Author                   = {J.~P.~Perdew and K.~Burke and M.~Ernzerhof},
  Journal                  = {Phys. Rev. Lett.},
  Year                     = {1996},
  Pages                    = {3865},
  Volume                   = {77},
  Doi                      = {10.1103/PhysRevLett.77.3865}
}

@Article{Pfaffenzeller1995,
  Title                    = {Miscibility of Hydrogen and Helium under Astrophysical Conditions},
  Author                   = {Pfaffenzeller, O. and Hohl, D. and Ballone, P.},
  Journal                  = {Phys. Rev. Lett.},
  Year                     = {1995},
  Pages                    = {2599--2602},
  Volume                   = {74},
  Doi                      = {10.1103/PhysRevLett.74.2599}
}

@article{He2010,
title = {First-principle study of solid neon under high compression},
journal = {Physica B: Condensed Matter},
volume = {405},
number = {20},
pages = {4335-4338},
year = {2010},
issn = {0921-4526},
doi = {https://doi.org/10.1016/j.physb.2010.07.037},
url = {https://www.sciencedirect.com/science/article/pii/S0921452610007362},
author = {Yi-guang He and Xiu-zhang Tang and Yi-kang Pu}
}

@article{Wang2024,
    author = {Wang, Zhao-Qi and Gu, Yun-Jun and Tang, Jun and Yan, Zheng-Xin and Xie, You and Wang, Yi-Xian and Chen, Xiang-Rong and Chen, Qi-Feng},
    title = "{Ab initio determination of melting and sound velocity of neon up to the deep interior of the Earth}",
    journal = {The Journal of Chemical Physics},
    volume = {160},
    number = {20},
    pages = {204711},
    year = {2024},
    month = {05},
    issn = {0021-9606},
    doi = {10.1063/5.0200412},
    url = {https://doi.org/10.1063/5.0200412},
    eprint = {https://pubs.aip.org/aip/jcp/article-pdf/doi/10.1063/5.0200412/19966637/204711\_1\_5.0200412.pdf},
}

@article{Tang2017,
    author = {Tang, J. and Chen, Q. F. and Fu, Z. J. and Li, Z. G. and Quan, W. L. and Gu, Y. J. and Zheng, J.},
    title = "{First-principles study of conducting behavior of warm dense neon}",
    journal = {Physics of Plasmas},
    volume = {24},
    number = {8},
    pages = {082709},
    year = {2017},
    month = {08},
    issn = {1070-664X},
    doi = {10.1063/1.5000526},
    url = {https://doi.org/10.1063/1.5000526},
    eprint = {https://pubs.aip.org/aip/pop/article-pdf/doi/10.1063/1.5000526/15648464/082709\_1\_online.pdf},
}

@article{Nghia2022,
title = {Equation-of-state and melting curve of solid neon and argon up to 100 GPa},
journal = {Vacuum},
volume = {196},
pages = {110725},
year = {2022},
issn = {0042-207X},
doi = {https://doi.org/10.1016/j.vacuum.2021.110725},
url = {https://www.sciencedirect.com/science/article/pii/S0042207X21006709},
author = {Nguyen {Van Nghia} and Ho Khac Hieu and Duong Dai Phuong}
}

@Article{Puestow2016,
  Title                    = {H/He demixing and the cooling behavior of Saturn},
  Author                   = {Robert Püstow and Nadine Nettelmann and Winfried Lorenzen and Ronald Redmer},
  Journal                  = {Icarus},
  Year                     = {2016},
  Pages                    = {323 - 333},
  Volume                   = {267},

  Doi                      = {https://doi.org/10.1016/j.icarus.2015.12.009},
  ISSN                     = {0019-1035},
  Url                      = {http://www.sciencedirect.com/science/article/pii/S0019103515005606}
}

@Article{Ross1983,
  Title                    = {The Equation Of State Of Molecular-Hydrogen At Very High-Density},
  Author                   = {Ross, M. and Ree, F. H. and Young, D. A.},
  Journal                  = {J. Chem. Phys.},
  Year                     = {1983},
  Number                   = {3},
  Pages                    = {1487--1494},
  Volume                   = {79},
  Doi                      = {10.1063/1.445939}
}

@Article{Scandolo2003,
  Title                    = {Liquid-liquid phase transition in compressed hydrogen from first-principles simulations},
  Author                   = {Scandolo, S},
  Journal                  = {Proc. Natl. Acad. Sci. USA},
  Year                     = {2003},
  Number                   = {6},
  Pages                    = {3051-3053},
  Volume                   = {100},
  Doi                      = {10.1073/pnas.0038012100}
}

@article{Bailey2021,
  title={Thermodynamically Governed Interior Models of Uranus and Neptune},
  author={Bailey, Elizabeth and Stevenson, David J.},
  journal={Planet. Sci. J.},
  volume={2},
  number={2},
  pages={64},
  year={2021},
  doi={10.3847/PSJ/abd1e0},
  publisher={American Astronomical Society}
}

@article{Scheibe2021,
  author = {Scheibe, Ludwig and Nettelmann, Nadine and Redmer, Ronald},
  title = {Thermal evolution of Uranus and Neptune - II. Deep thermal boundary layer},
  DOI= {10.1051/0004-6361/202140663},
  url= {https://doi.org/10.1051/0004-6361/202140663},
  journal = {Astron. Astrophys.},
  year = {2021},
  volume = {650},
  pages = {A200},
}

@Article{Schoettler2018,
  author    = {Sch\"ottler, Manuel and Redmer, Ronald},
  title     = {Ab Initio Calculation of the Miscibility Diagram for Hydrogen-Helium Mixtures},
  journal   = {Phys. Rev. Lett.},
  year      = {2018},
  volume    = {120},
  pages     = {115703},
  month     = {Mar},
  doi       = {10.1103/PhysRevLett.120.115703},
  issue     = {11},
  numpages  = {6},
  publisher = {American Physical Society},
  url       = {https://link.aps.org/doi/10.1103/PhysRevLett.120.115703},
}

@article{Soubiran2015b,
doi = {10.1088/0004-637X/806/2/228},
url = {https://dx.doi.org/10.1088/0004-637X/806/2/228},
year = {2015},
month = {jun},
publisher = {The American Astronomical Society},
volume = {806},
number = {2},
pages = {228},
author = {François Soubiran and Burkhard Militzer},
title = {MISCIBILITY CALCULATIONS FOR WATER AND HYDROGEN IN GIANT PLANETS},
journal = {Astrophys. J.},
}

@Article{Stevenson1975,
  Title                    = {Thermodynamics and phase separation of dense fully ionized hydrogen-helium fluid mixtures},
  Author                   = {D.~J. Stevenson},
  Journal                  = {Phys. Rev. B},
  Year                     = {1975},

  Month                    = {Nov},
  Number                   = {10},
  Pages                    = {3999--4007},
  Volume                   = {12},
  Doi                      = {10.1103/PhysRevB.12.3999}
}

@Article{Stevenson1977a,
  Title                    = {The phase diagram and transport properties for hydrogen-helium fluid planets},
  Author                   = {D.~J. Stevenson and E.~E. Salpeter},
  Journal                  = {Astrophys. J. Suppl. Ser.},
  Year                     = {1977},
  Pages                    = {221-237},
  Volume                   = {35},
  Doi                      = {10.1086/190478}
}

@Article{Vorberger2007,
  Title                    = {Hydrogen-helium mixtures in the interiors of giant planets},
  Author                   = {J.~Vorberger and I.~Tamblyn and B.~Militzer and S.~A.~Bonev},
  Journal                  = {Phys. Rev. B},
  Year                     = {2007},
  Pages                    = {024206},
  Volume                   = {75},
  Doi                      = {10.1103/PhysRevB.75.024206}
}

@Article{Vos1991,
  Title                    = {High pressure phase diagram of helium-hydrogen calculated through fluid integral equations and density functional theory of freezing},
  Author                   = {W L Vos and A de Kuijper and J L Barrat and J A Schouten},
  Journal                  = {J. Phys.: Condens. Matter},
  Year                     = {1991},
  Pages                    = {1613},
  Volume                   = {3},
  Doi                      = {10.1088/0953-8984/3/11/019}
}

@Article{Wilson2010,
  author    = {Wilson, Hugh F. and Militzer, Burkhard},
  title     = {Sequestration of Noble Gases in Giant Planet Interiors},
  journal   = {Phys. Rev. Lett.},
  year      = {2010},
  volume    = {104},
  pages     = {121101},
  month     = {Mar},
  doi       = {10.1103/PhysRevLett.104.121101},
  issue     = {12},
  numpages  = {4},
  publisher = {American Physical Society},
  url       = {https://link.aps.org/doi/10.1103/PhysRevLett.104.121101},
}

@Article{Niemann1996,
  author    = {Niemann, Hasso B. and Atreya, Sushil K. and Carignan, George R. and Donahue, Thomas M. and Haberman, John A. and Harpold, Dan N. and Hartle, Richard E. and Hunten, Donald M. and Kasprzak, Wayne T. and Mahaffy, Paul R. and Owen, Tobias C. and Spencer, Nelson W. and Way, Stanley H.},
  title     = {The Galileo Probe Mass Spectrometer: Composition of Jupiter{\textquoteright}s Atmosphere},
  journal   = {Science},
  year      = {1996},
  volume    = {272},
  number    = {5263},
  pages     = {846--849},
  issn      = {0036-8075},
  doi       = {10.1126/science.272.5263.846},
  publisher = {American Association for the Advancement of Science},
  url       = {https://science.sciencemag.org/content/272/5263/846},
}

@Article{Helled2020a,
  author    = {Helled, Ravit and Nettelmann, Nadine and Guillot, Tristan},
  title     = {Uranus and Neptune: Origin, Evolution and Internal Structure},
  journal   = {Space Sci. Rev.},
  year      = {2020},
  volume    = {216},
  number    = {3},
  month     = {Mar},
  issn      = {1572-9672},
  publisher = {Springer Science and Business Media LLC},
  url       = {http://dx.doi.org/10.1007/s11214-020-00660-3},
}

@Article{Nettelmann2016,
  author   = {N. Nettelmann and K. Wang and J.J. Fortney and S. Hamel and S. Yellamilli and M. Bethkenhagen and R. Redmer},
  title    = {Uranus evolution models with simple thermal boundary layers},
  journal  = {Icarus},
  year     = {2016},
  volume   = {275},
  pages    = {107-116},
  issn     = {0019-1035},
  doi      = {https://doi.org/10.1016/j.icarus.2016.04.008},
  url      = {https://www.sciencedirect.com/science/article/pii/S0019103516300537},
}

@Article{Zaghoo2016,
  author    = {Zaghoo, Mohamed and Salamat, Ashkan and Silvera, Isaac F.},
  title     = {Evidence of a first-order phase transition to metallic hydrogen},
  journal   = {Phys. Rev. B},
  year      = {2016},
  volume    = {93},
  pages     = {155128},
  month     = {Apr},
  doi       = {10.1103/PhysRevB.93.155128},
  issue     = {15},
  numpages  = {7},
  publisher = {American Physical Society},
  url       = {https://link.aps.org/doi/10.1103/PhysRevB.93.155128},
}

@Article{Dzyabura2013,
  author    = {Dzyabura, Vasily and Zaghoo, Mohamed and Silvera, Isaac F.},
  title     = {Evidence of a liquid{\textendash}liquid phase transition in hot dense hydrogen},
  journal   = {Proc. Natl. Acad. Sci. USA},
  year      = {2013},
  volume    = {110},
  number    = {20},
  pages     = {8040--8044},
  issn      = {0027-8424},
  doi       = {10.1073/pnas.1300718110},
  publisher = {National Academy of Sciences},
  url       = {https://www.pnas.org/content/110/20/8040},
}

@Article{Ohta2015,
  author    = {Ohta, Kenji and Ichimaru, Kota and Einaga, Mari and Kawaguchi, Sho and Shimizu, Katsuya and Matsuoka, Takahiro and Hirao, Naohisa and Ohishi, Yasuo},
  title     = {Phase boundary of hot dense fluid hydrogen},
  journal   = {Sci. Rep.},
  year      = {2015},
  volume    = {5},
  number    = {1},
  pages     = {1--7},
  doi       = {10.1038/srep16560},
  publisher = {Nature Publishing Group},
}

@Article{Ross1996,
  author    = {Ross, Marvin},
  title     = {Insulator-metal transition of fluid molecular hydrogen},
  journal   = {Phys. Rev. B},
  year      = {1996},
  volume    = {54},
  pages     = {R9589--R9591},
  month     = {Oct},
  doi       = {10.1103/PhysRevB.54.R9589},
  issue     = {14},
  numpages  = {0},
  publisher = {American Physical Society},
  url       = {https://link.aps.org/doi/10.1103/PhysRevB.54.R9589},
}
\end{document}